\documentclass[apjl]{emulateapj}
\usepackage{amsmath}
\usepackage{color}

\shortauthors{Kreidberg et al.}

\begin{document}

\title{A Detection of Water in the Transmission Spectrum of the Hot Jupiter WASP-12\lowercase{b} and Implications for its Atmospheric Composition}

\author{
Laura Kreidberg\altaffilmark{1,8},
Michael R. Line\altaffilmark{2},
Jacob L. Bean\altaffilmark{1,9},
Kevin B. Stevenson\altaffilmark{1,10},
Jean-Michel D{\'e}sert\altaffilmark{3},
Nikku Madhusudhan\altaffilmark{4},
Jonathan J. Fortney\altaffilmark{2},
Joanna K. Barstow\altaffilmark{5},
Gregory W. Henry\altaffilmark{6},
Michael H. Williamson\altaffilmark{6}, \&
Adam P. Showman\altaffilmark{7}
}

\email{E-mail: laura.kreidberg@uchicago.edu}

\altaffiltext{1}{Department of Astronomy and Astrophysics, University of Chicago, 5640 S.~Ellis Ave, Chicago, IL 60637, USA}
\altaffiltext{2}{Department of Astronomy and Astrophysics, University of California,Santa Cruz, CA 95064, USA}
\altaffiltext{3}{CASA, Department of Astrophysical \& Planetary Sciences, University of Colorado, 389-UCB, Boulder, CO 80309, USA}
\altaffiltext{4}{Institute for Astronomy, University of Cambridge, Cambridge CB3 OHA, UK}
\altaffiltext{5}{Department of Physics, University of Oxford, Oxford OX1 3RH, UK}
\altaffiltext{6}{Center of Excellence in Information Systems, Tennessee State University, Nashville, TN 37209, USA}
\altaffiltext{7}{Department of Planetary Sciences and Lunar and Planetary Laboratory, The University of Arizona, Tuscon, AZ 85721, USA}
\altaffiltext{8}{National Science Foundation Graduate Research Fellow}
\altaffiltext{9}{Packard Fellow}
\altaffiltext{10}{Sagan Fellow}

\begin{abstract}
Detailed characterization of exoplanets has begun to yield measurements of their atmospheric properties that constrain the planets' origins and evolution. For example, past observations of the dayside emission spectrum of the hot Jupiter WASP-12b indicated that its atmosphere has a high carbon-to-oxygen ratio (C/O $>$ 1), suggesting it had a different formation pathway than is commonly assumed for giant planets.  Here we report a precise near-infrared transmission spectrum for WASP-12b based on six transit observations with the \textit{Hubble Space Telescope}/Wide Field Camera 3.  We bin the data in 13 spectrophotometric light curves from 0.84 - 1.67 $\mu$m and measure the transit depths to a median precision of 51 ppm.  We retrieve the atmospheric properties using the transmission spectrum and find strong evidence for water absorption (7\,$\sigma$ confidence).  This detection marks the first high-confidence, spectroscopic identification of a molecule in the atmosphere of WASP-12b.   The retrieved 1\,$\sigma$ water volume mixing ratio is between $10^{-5}-10^{-2}$, which is consistent with C/O $>$ 1 to within 2\,$\sigma$.  However, we also introduce a new retrieval parameterization that fits for C/O and metallicity under the assumption of chemical equilibrium.  With this approach, we constrain C/O to $0.5^{+0.2}_{-0.3}$ at $1\,\sigma$ and rule out a carbon-rich atmosphere composition (C/O\,$>1$) at $>3\,\sigma$ confidence. Further observations and modeling of the planet's global thermal structure and dynamics would aid in resolving the tension between our inferred C/O and previous constraints.  Our findings highlight the importance of obtaining high-precision data with multiple observing techniques in order to obtain robust constraints on the chemistry and physics of exoplanet atmospheres. 
\end{abstract}

\keywords{planets and satellites: atmospheres --- planets and satellites:  composition --- planets and satellites: individual: WASP-12b}

\section{Introduction}
The chemical composition of a planetary atmosphere provides a rich record of the planet's formation conditions and evolutionary history.  Measurements of the composition can constrain the planet's formation mechanism, its formation location in the protoplanetary disk, the surface density and composition of planetesimals at the formation site, the relative accretion rates of gas and solids, and possible migration pathways \citep[e.g.][]{atreya99, owen99, gautier01, atreya03, hersant04, lodders04, dodsonrobinson09, oberg11, madhusudhan11b, madhusudhan14b}.  

Because there are many factors that influence atmospheric chemical composition, a large sample size of planets is required to develop a comprehensive theory of giant planet formation.  Fortunately, the sample of known extrasolar giant planets is large and growing, and recent observations of these planets have begun to yield basic constraints on their atmospheric chemistry. These include inferences of the carbon-to-oxygen ratio (C/O) and absolute water abundance, which in some cases rival our knowledge of those quantities for the Solar System planets \citep[e.g.][]{madhusudhan11a, konopacky13, line14, kreidberg14b, brogi14, madhusudhan14a}.  

One of the best studied exoplanet atmospheres is that of the transiting hot Jupiter WASP-12b.  This planet is a $1.4\, M_\mathrm{Jup}$, $1.8\, R_\mathrm{Jup}$ gas giant orbiting a late-F host star with a period of just 1.1 days \citep{hebb09}.  The brightness of the host star (H=10.2), the planet's high equilibrium temperature (2500 K), and the planet's large size make WASP-12b a favorable target for atmosphere characterization.  The system has been observed extensively from the ground and space to measure the planet's transmission spectrum, dayside emission spectrum, and thermal phase variation \citep{lopez-morales10, fossati10, campo11, croll11, cowan12, crossfield12, haswell12, zhao12, fohring13, sing13, swain13, mandell13,copperwheat13, stevenson14a, stevenson14b, burton15, nichols15, croll15}.

An intriguing possibility has emerged from these observations of WASP-12b, which is that the planet has a carbon-rich atmospheric composition \citep[C/O $> 1$, compared to the solar value of 0.55;][]{asplund09}.  The high C/O interpretation was first suggested by \cite{madhusudhan11a} as the best explanation for the planet's dayside emission spectrum.  This result was contested by subsequent work \citep{crossfield12, line14}; however, the most recent comprehensive analysis of the dayside spectrum reaffirmed the inference of high C/O \citep{stevenson14b}.

The claim of high C/O in WASP-12b's atmosphere has motivated substantial theoretical and observational work. This includes attempts to measure C/O for additional planets \citep[e.g.][]{brogi14, line14}, studies of the effect of C/O on atmospheric chemistry \citep{madhusudhan11b, madhusudhan12, kopparapu12, moses13, venot15}, inferences of C/O in exoplanet host stars \citep{teske13, teske14}, and predictions of C/O from planet formation theory \citep{madhusudhan11b, oberg11, madhusudhan14b, alidib14}. 

We note, however, that the evidence for a carbon-rich atmosphere on WASP-12b is not definitive.  The high C/O inference is based primarily on the photometric secondary eclipse depth from \textit{Spitzer} at 4.5 $\mu$m.  The low brightness temperature for the planet in this bandpass is best explained by absorption from carbon-bearing species \citep[either CO, HCN, or C$_2$H$_2$;][]{madhusudhan11a, kopparapu12, moses13, stevenson14a}, but photometry alone cannot uniquely identify which molecule is the dominant absorber.  No other molecular features have been confidently identified in the emission spectrum, and measurements of the planet's transmission spectrum have yielded even less conclusive constraints on the C/O \citep{sing13, mandell13, swain13, stevenson14a, madhusudhan14a}.  High-precision spectroscopy is thus needed to obtain unambiguous determination of the atmospheric composition of this important planet.

In this work, we report a new, precise measurement of WASP-12b's transmission spectrum over the wavelength range 0.84 to 1.67 $\mu$m.  The outline of the paper is as follows.  In \S\,\ref{sec:obs_data} we present the observations and data reduction.  The light curves fits and measurement of the transmission spectrum are outlined in \S\,\ref{sec:lc}.  We compare our results to previous measurements in \S\,\ref{sec:comparison}. We describe a retrieval of the planet's atmospheric properties based on this transmission spectrum in \S\,\ref{sec:retrieval} and discuss implications for the chemical composition in \S\,\ref{sec:ctoo}. We conclude in \S\,\ref{sec:conclusion}.

\section{Observations and Data Reduction}
\label{sec:obs_data}
\subsection{Observations}
\label{subsec:observations}
\begin{figure}
\resizebox{\hsize}{!}{\includegraphics{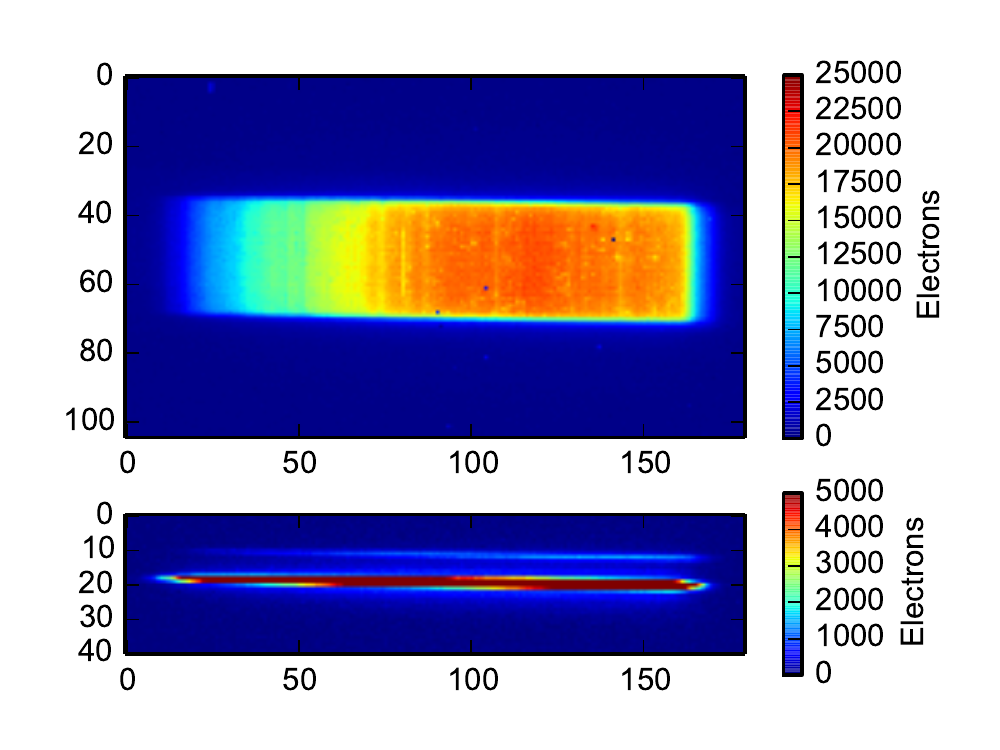}}
\caption{Raw \textit{HST}/WFC3 images taken with the G102 grism. Spatial scan and staring mode data are shown in the top and bottom panels, respectively.  The images are cutouts from a 256x256 pixel subarray.  The spectrum of WASP-12A's binary companion is visible in the bottom panel near row 10.}
\label{fig:rawdata}
\end{figure}

We obtained time series spectroscopy during six transits of WASP-12b between UT 1 January and 4 March 2014 using the Wide Field Camera 3 (WFC3) IR detector on the \textit{Hubble Space Telescope} (\textit{HST}) as part of \textit{HST} GO Program 13467.  Three of the transit observations used the G102 grism, which provides spectral coverage from $0.82-1.12$ $\mu$m, and the other three used the G141 grism, which spans the range $1.12-1.65$ $\mu$m.  The G141 grism has been widely used for exoplanet transit spectroscopy, but these observations are the first to use the G102 grism for this purpose.  The additional wavelength coverage from the G102 grism provides access to features from more molecules and absorption bands, giving us greater leverage in constraining the atmospheric composition of the planet. Each transit observation (called a visit) consisted of five consecutive, 96-minute \textit{HST} orbits.  WASP-12 was visible during approximately 45 minutes per orbit and occulted by the Earth for the remainder of the time.  We took a direct image of the target with the F126N narrow-band filter at the beginning of each orbit for wavelength calibration.  Example staring mode and spatial scan images are shown in Figure \ref{fig:rawdata}.

The spectroscopic data were obtained in spatial scan mode with the 256x256 subarray, using the SPARS10, NSAMP=16 readout pattern,  which has an exposure time of 103.1\,s.  The scan rates were 0.04 and 0.05 arcsec/second for the G102 and G141 grisms, respectively.  The spectra extend roughly 40 pixels in the spatial direction, with peak per-pixel counts below 25,000 electrons for both grisms.  We alternated between forward and reverse scanning along the detector to decrease instrumental overhead time. This setup yielded 19 exposures per orbit and a duty cycle of 74\%. 

In addition to the spatial scan data, we also obtained 10 staring mode spectra in each grism during the first orbit of the first visit.  WASP-12 is a triple star system: WASP-12\,A hosts the planet, and WASP-12\,BC is an M-dwarf binary separated from WASP-12\,A by about 1" \citep{bergfors13,bechter14}.  We use these staring mode data to resolve the spectrum of WASP-12\,A from WASP-12\,BC, enabling us to correct for dilution to the planet's transit light curve due to the binary.  The detector orientation was set to $178.7^\circ$ in order to spatially separate the spectra.  At this orientation, the spectra are separated by 7.9 pixels (compared to the FWHM of 1.1 pixel at 1.4 $\mu$m).  We describe the dilution correction in detail in \S\,\ref{subsec:depth_meas}. 

\begin{figure}
\resizebox{\hsize}{!}{\includegraphics{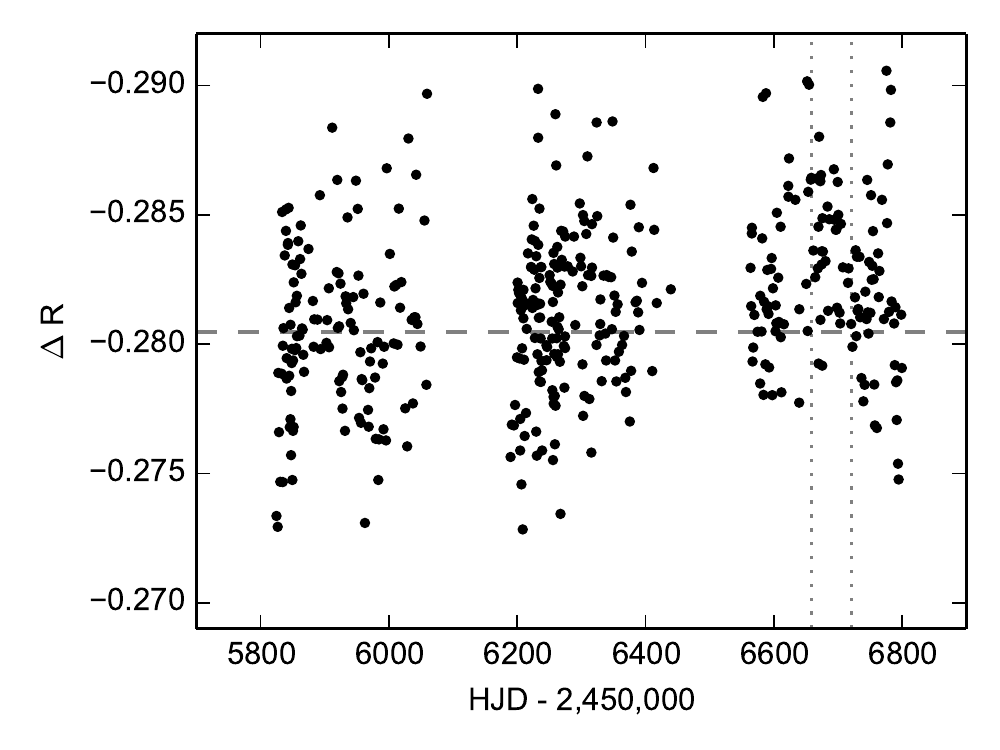}}
\caption{Nightly Cousins $R$-band photometric observations of WASP-12 over three observing seasons (points).  The data are differential magnitudes relative to the average brightness of 17 comparison stars.  The differential magnitudes have a mean of -0.281 and a standard deviation of 0.003. The horizontal dashed line indicates the mean brightness from the first observing season.  The vertical dotted lines span the time range of the HST observations.}
\label{fig:monitoring}
\end{figure}

\begin{deluxetable*}{ccccc}
\tabletypesize{\scriptsize}
\tablecolumns{5}
\tablewidth{0pc}
\tablecaption{Summary of Photometric Observations for WASP-12}
\tablehead{ \colhead{Season} & \colhead{$N_{obs}$} & \colhead{Date Range} & \colhead{Sigma} & \colhead{Seasonal Mean} \\
& & \colhead{(HJD - 2,400,000)} & (mag) & (mag)}
\startdata
2011-2012 & 126 & 55824-56059 & 0.00325 & $-0.28048\pm0.00029$ \\
2012-2013 & 169 & 56189-56439 & 0.00297 & $-0.28119\pm0.00023$ \\
2013-2014 & 133 & 56463-56799 & 0.00316 & $-0.28253\pm0.00027$
\enddata
\label{tab:photometry}
\end{deluxetable*}

We also obtained photometric monitoring of the WASP-12 system to search for stellar activity.  We acquired 428 out-of-transit R-band images over the 2011-2012, 2012-2013, and 2013-2014 observing seasons with Tennessee State University's Celestron 14-inch automated imaging telescope.  We tested for variability from star spots by calculating differential magnitudes for WASP-12 relative to 17 comparison stars (shown in Figure\,\ref{fig:monitoring}).  The standard deviation of the differential magnitudes is 0.3\%, which is comparable to the photon noise for the data.  The brightness of the star increases by approximately 0.001 mag per year. There are no significant periodicities between 1 and 200 days.  To calculate the impact of star spots on our transmission spectrum, we took 0.3\% as an upper limit for the variability of WASP-12.  Variability of this amplitude could be produced by star spots 300 K cooler than the star's effective temperature \citep[6300 K;][]{hebb09}, covering 3\% of the photosphere.  Based on the formalism outlined in \cite{berta11} and \cite{desert11}, we calculated that the maximum variation in transit depth due to spots with these properties is of order $10^{-5}$, which is below the precision of our transit depth measurements. 

\begin{figure}
\resizebox{\hsize}{!}{\includegraphics{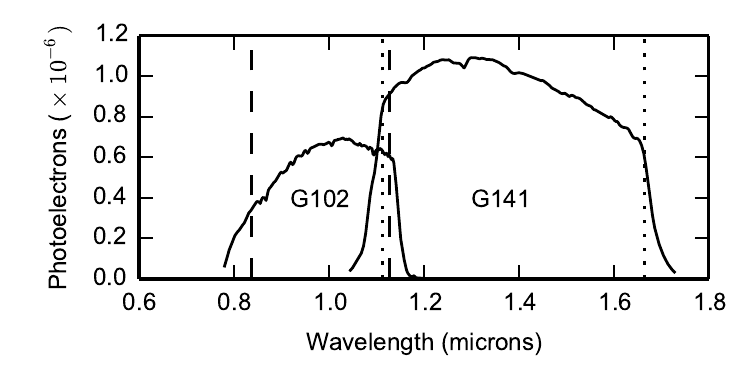}}
\caption{Example extracted spectra (solid lines).  The wavelength ranges covered by the transmission spectra are indicated with dashed and dotted lines (for the G102 and G141 data, respectively).  The uncertainties on the spectra are smaller than the plot linewidth.}
\label{fig:spectra}
\end{figure}

\subsection{Data Reduction}
We reduced the \textit{HST}/WFC3 data using the custom pipeline described in \cite{kreidberg14a}.  Example extracted spectra are shown in Figure\,\ref{fig:spectra}.  We bin the spectra into 24- and 15-pixel wide channels (for G102 and G141, respectively) to obtain a total of 13 spectrophotometric light curves at resolution $R \equiv \lambda/\Delta \lambda = 15-25$.  We also sum the spectra over the full wavelength range to create broadband (``white") light curves for each grism. The uncertainty in the flux per exposure is determined by adding in quadrature the photon noise, the read noise (22 electrons per differenced image), and the error in the estimate of the background, which we determine from the median absolute deviation of the flux values for the background pixels. 
 
We also extract spectra from the staring mode data to determine the wavelength-dependent flux ratio of WASP-12\,BC to WASP-12\,A.  The spectra are spatially separated on the detector, so to extract them we fit a double Gaussian model to each column of the final detector read.  We obtain final estimates and uncertainties for the spectra of WASP-12\,A and WASP-12\,BC by taking the mean and standard deviation of the spectra from all 10 exposures. These extracted spectra are each assigned a wavelength solution using the centroids of the stars in the direct image.

\section{Light Curve Fits}
\label{sec:lc}
\subsection{Broadband Light Curves}
We fit the broadband light curves with a transit model \citep{mandel02} and an analytic function to correct for instrument systematics.  Transit light curve observations with \textit{HST}/WFC3 have several well-documented systematic trends in flux with time, including visit-long slopes and orbit-long exponential ramps \citep{berta12,swain13,kreidberg14a, stevenson14d}.  The first orbit of a visit has a larger amplitude ramp than subsequent orbits, so we follow established practice and do not use this initial orbit in the light curve fits.  We also discard the first exposure from the remaining orbits, which improves the fit quality.  We fit the remaining data (216 exposures for each grism) with a systematics model based on the \texttt{model-ramp} parameterization of \cite{berta12}.  The model $M(t)$ has the form:

\begin{align}
 M(t) &= T(t)\times(c\,S(t) + v\,t_\mathrm{v})	 \nonumber \\
	&\qquad {} \times (1 - \exp(-a\,t_\mathrm{orb} - b - D(t))).
\end{align}

In our fits, $T(t)$ is the transit model, or relative stellar flux as a function of time $t$ (in BJD$_\mathrm{TDB}$).  The free parameters for the transit model are the planet-to-star radius ratio $k$, a linear limb darkening parameter $u$, the ratio of semi-major axis to stellar radius $a/R_s$, the orbital inclination $i$, and the time of mid-transit $T_0$. The light curves have poor coverage of the planet's ingress, so we put priors on $a/R_s$ and $i$ to enable the measurement of $T_0$.  We use Gaussian priors with mean and standard deviation $2.91 \pm 0.02$ and $80.56 \pm 0.03^\circ$ (for $a/R_s$ and $i$, respectively), based on estimates of those parameters from \cite{stevenson14a}.  We use an orbital period $P = 1.091424$ days and assume a circular orbit \citep{campo11}.  The data from the two grisms are fit separately, but for each grism the three transits are fit simultaneously.  We fit unique values of $k$ and $T_0$ to each transit, but tie the values for $u$, $a/R_s$, and $i$ over all the transits.  

We fit the instrument systematics with a constant normalization term $c$, a scaling factor $S(t)$, a visit-long linear slope $v$, and an orbit-long exponential ramp with rate constant $a$, amplitude $b$, and delay $D(t)$.  The timescale $t_\mathrm{v}$ corresponds to time relative to the expected transit midpoint for each visit, and $t_\mathrm{orb}$ is time since the first exposure in an orbit (both in BJD$_\mathrm{TDB}$).  The scaling factor $S(t)$ is equal to $1$ for exposures with forward spatial scanning and $s$ for reverse scanning; this accounts for a small offset in normalization between the scan directions caused by the upstream-downstream effect of the detector readout \citep{mccullough12}.  The function $D(t)$ is equal to $d$ for times $t$ during the first fitted orbit and $0$ elsewhere.  A negative value for $d$ implies that the ramp amplitude is larger in the first orbit than in subsequent orbits.  The parameters $u$, $a$, $b$, and $d$ were constrained to the same value for all the transits, whereas $k$, $s$, $c$, and $v$ were allowed to vary between transits.  

There are a total of 21 free parameters in the fits to each grism's broadband light curve.  We estimated the parameters and their uncertainties with a Markov chain Monte Carlo (MCMC) fit to the data, using the \texttt{emcee} package for Python \citep{foremanmackey13}.  For the best-fit light curves, we calculated the Durbin-Watson statistic to test for time-correlated noise.  The values were 1.77 and 2.17 for the G102 and G141 white light curves, indicating that the residuals are uncorrelated at the 1\% significance level.  The reduced chi-squared values ($\chi^2_\nu$) for the best-fit light curves are 1.45 and 1.34, and the residuals are 164 and 115 ppm (37 and 25\% larger than the predicted photon+read noise) for the G102 and G141 data, respectively.  

\begin{deluxetable}{cccc}
\tabletypesize{\scriptsize}
\tablecolumns{3}
\tablewidth{0pc}
\tablecaption{Transit Times}
\tablehead{ 
\colhead{Observation Start} & \colhead{$T_c$\tablenotemark{a}} & \colhead{Uncertainty} & \\
\colhead{(UT)} & \colhead{(BJD$_\mathrm{TDB}$)} & \colhead{(1\,$\sigma$)}} \\
\startdata
Jan 01 2014 & 6659.07598 & 3.4E-4 \\
Jan 16 2014 & 6674.35560 & 2.8E-4 \\
Feb 05 2014 & 6694.00161 & 2.9E-4 \\
Feb 15 2014 & 6703.82417 & 2.9E-4 \\
Mar 02 2014 & 6719.10428 & 3.4E-4 \\
Mar 04 2014 & 6721.28692 & 3.4E-4 
\enddata
\tablenotetext{a}{We report the time of central transit $T_c$ in BJD$_\mathrm{TDB}$ - 2,450,000.}
\label{tab:transittimes}
\end{deluxetable}		

To obtain conservative estimates of the errors for the white light curve transit parameters, we redid the MCMC analyses with the per-exposure uncertainties scaled by a constant factor (1.20 and 1.16) chosen to yield $\chi^2_\nu = 1$.  We report the transit times from the white light curve fits in Table\,\ref{tab:transittimes}.   These values extend the baseline of precise transit times by two years from the \cite{stevenson14a} measurements and will aid in testing the possibility of perihelion precession for this system \citep[first studied by][]{campo11}.   The white light curves are consistent with the priors on $a/R_s$ and $i$ but do not yield improved values for those parameters due to the poor phase coverage of the transit ingress. 

\subsection{Spectroscopic Light Curves}
\label{subsec:speclc}

\begin{figure}[ht!]
\resizebox{\hsize}{!}{\includegraphics{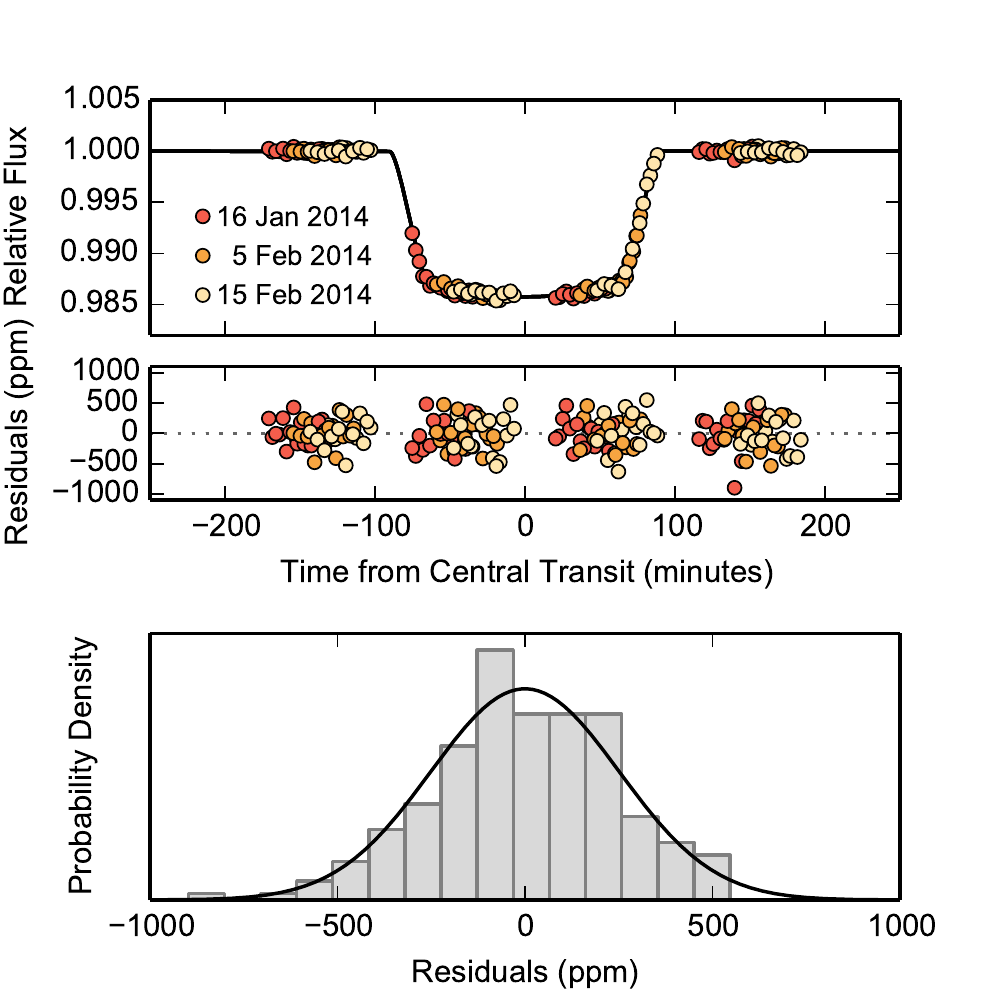}}
\caption{An example light curve fit to the $1.320 - 1.389$ $\mu$m spectroscopic channel from the G141 grism.  The top panel shows the best fitting model light curve (black line), overlaid with the systematics-corrected data (points).  Residuals from the light curve fit are shown in the middle panel.  The bottom panel shows a normalized histogram of the residuals compared to a Gaussian probability density with mean zero and standard deviation equal to the predicted photon+read noise (252 ppm).}
\label{fig:lc}
\end{figure}

We fit the spectroscopic light curves with a similar model as we use for the broadband data.  The only differences are that we hold $a/R_s$ and $i$ fixed to the prior mean values and fix the mid-transit times to the white light curve best fit values.  We also solve for a constant rescaling parameter for the photometric uncertainties in each spectroscopic channel to ensure that the reduced $\chi^2$ for the light curve fits is unity.  We chose to fix $a/R_s$, $i$, and $T_c$ because they impact the mean spectroscopic transit depth only, not the relative depths. Changes in the mean transit depth do not significantly affect our retrieval results, as the planet-to-star radius ratio is included as a free parameter in the atmospheric retrieval.

\begin{figure*}[h!]
\resizebox{1.0\textwidth}{!}{\includegraphics{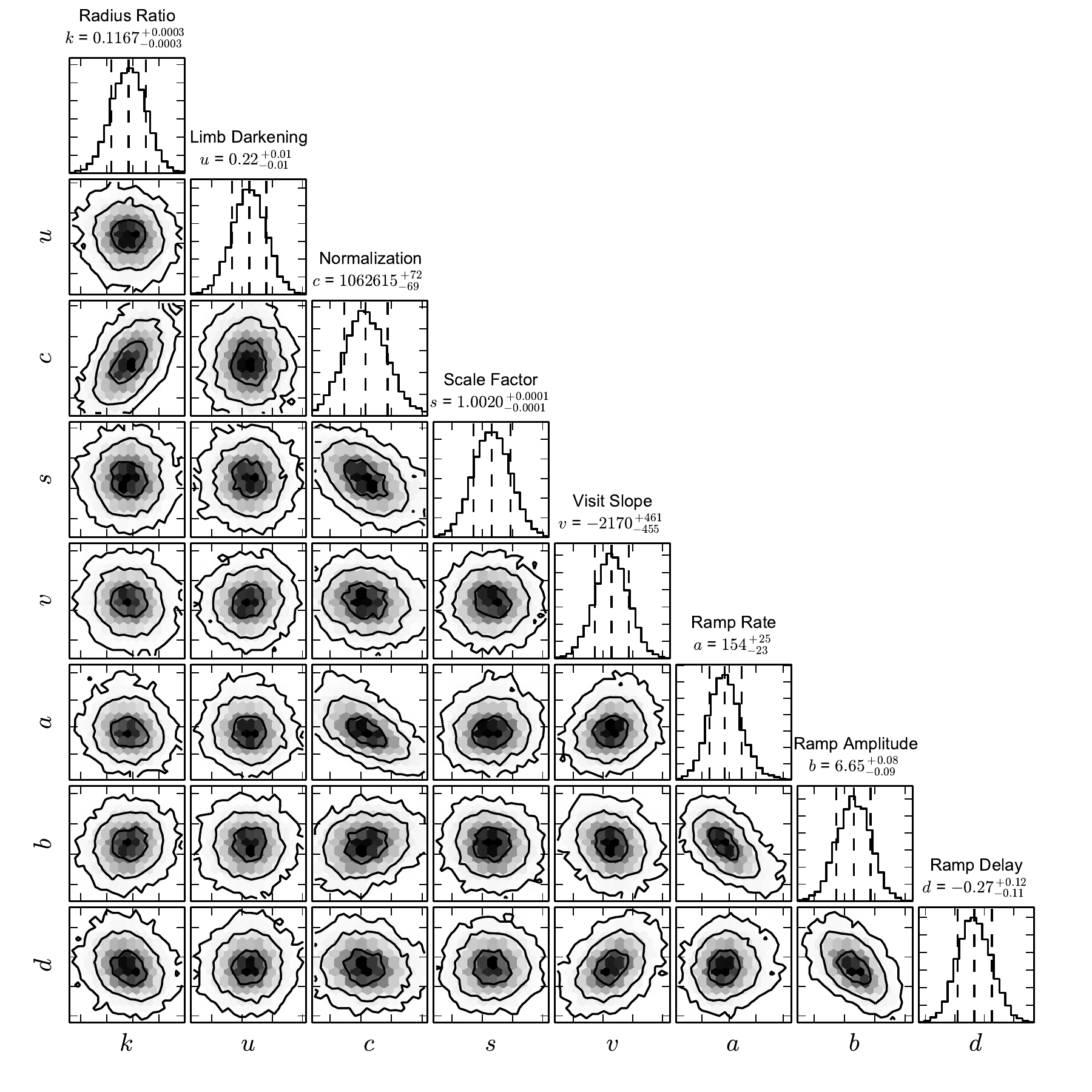}}
\caption{A pairs plot showing distributions of MCMC fit parameters for the $1.320-1.389$ $\mu$m light curve.  The off-diagonal panels show marginalized posterior probability for pairs of parameters, with 1, 2, and 3\,$\sigma$ credible intervals indicated with black contours.  The gray shading corresponds to probability density (darker for higher probability).  The panels on the diagonal show distributions of each parameter marginalized over the other model parameters, with the median and 68\% credible interval marked with dashed lines. The planet-to-star radius ratio, $k$, is not strongly correlated with any of the other fit parameters.  For parameters that are allowed to vary between transit observations ($k$, $c$, $s$, and $v$), we show distributions for the 16 Jan 2014 transit.}
\label{fig:pairslk}
\end{figure*}

We fit each of the spectroscopic channel light curves independently.  We achieve nearly photon-limited precision: the median rescaling factor for the photometric uncertainties is 1.1.  The light curves do not exhibit statistically significant time-correlated noise, based on a Durbin-Watson test at the 1\% significance level. An example spectroscopic light curve fit for the $1.320-1.389$ $\mu$m channel from the G141 grism is shown in Figure\,\ref{fig:lc}.  We show a pairs plot of the fit parameters for the same channel in Figure\,\ref{fig:pairslk}.  

We tested an analytic model for the systematics that included a quadratic term for the visit-long trend.  Previous analyses of WFC3 data have suggested that a quadratic model produces better light curve fits \citep{stevenson14d}. However, for this data set we find that the quadratic model is disfavored according to the Bayesian Information Criterion (BIC), with typical $\Delta$\,BIC values of 10 compared to the linear model.  In any case, the main consequence of adding a quadratic term is to shift the transmission spectrum up or down. This affects our estimate of the planetary radius, but the effect is small relative to the error introduced by uncertainties in the stellar radius.

We also explored modeling the instrument systematics with the \texttt{divide-white} technique, which has been applied successfully to several other data sets \citep{stevenson14a, kreidberg14a, knutson14b}.  This method assumes the systematics are independent of wavelength. For the WASP-12 data, however, this assumption is not appropriate.  WFC3 instrument systematics are known to depend on detector illumination \citep{berta12, swain13}, and in our data the mean pixel fluence varies by 30\% between the spectroscopic channels (see Figure\,\ref{fig:spectra}).

\begin{figure}
\resizebox{\hsize}{!}{\includegraphics{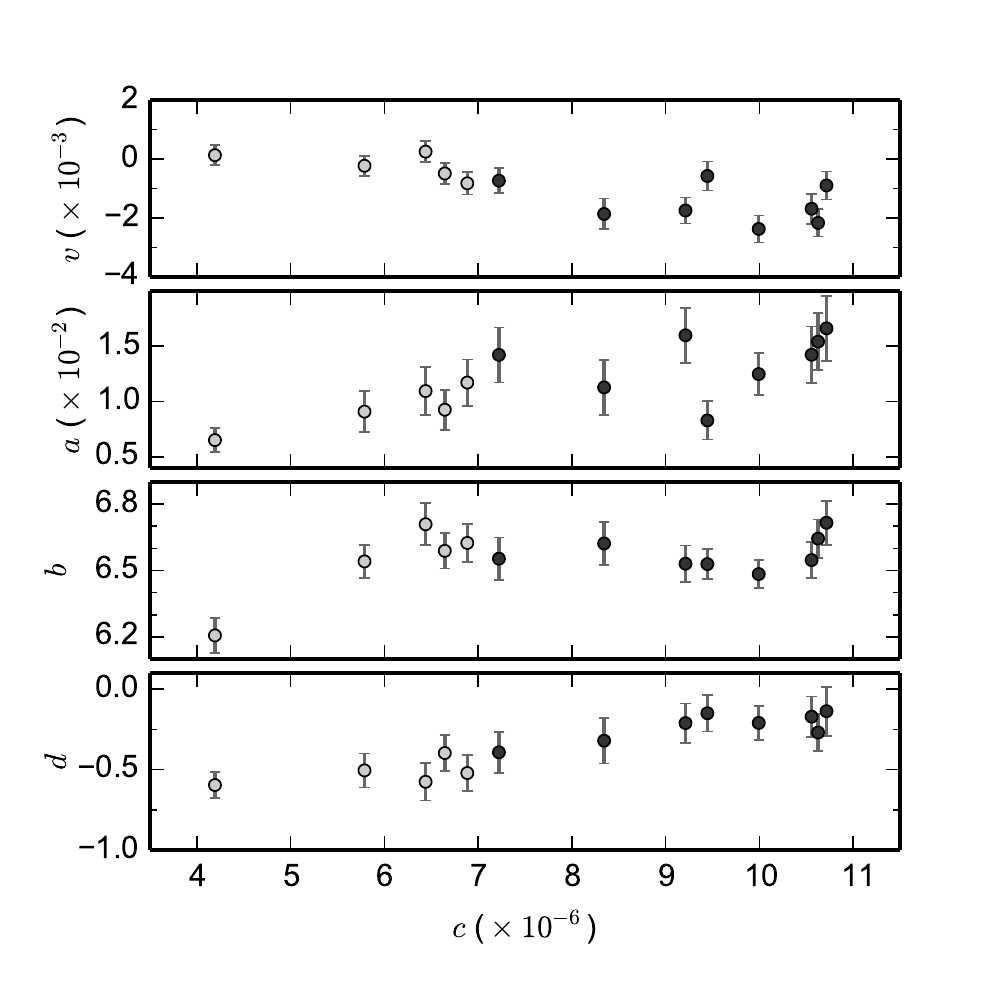}}
\caption{Systematics decorrelation parameters from the light curve fits plotted as a function of normalization $c$.  We show (from top to bottom) the visit-long slope $v$, ramp rate constant $a$, ramp amplitude $b$, and ramp delay $d$.  The normalization constant $c$ is linearly proportional to the per-pixel flux in a spectroscopic channel.  Measurements from the G102 and G141 data are shown in light and dark gray, respectively.  The error bars indicate 1\,$\sigma$ uncertainties from an MCMC fit to the light curves using the analytic model for instrument systematics.}
\label{fig:sys}
\end{figure}

To illustrate the dependence of the systematics on the illumination level, we show in Figure\,\ref{fig:sys} the systematics decorrelation parameters from the analytic model as a function of light curve normalization $c$.  The value $c$ represents the baseline flux level in each spectroscopic light curve. This value is directly proportional to the mean detector illumination in the channel.  We note several qualitative trends in the decorrelation parameters: with increasing illumination, the visit-long slope decreases, the delay term increases toward zero, and the ramp rate increases.  

The upshot of these trends is that there are residual systematics for the \texttt{divide-white} light curve fits that are correlated with detector illumination. We therefore only report the results for the analytic model.

\subsubsection{Limb Darkening Models}
We tested fixing the limb darkening to values predicted by stellar models.  We used theoretical quadratic limb darkening coefficients from both PHOENIX and Kurucz models \citep{hauschildt99, castelli04} generated for \cite{stevenson14a}.  These theoretical coefficients yielded lower quality light curve fits than we obtained from empirically estimating a linear limb darkening parameter.  For both PHOENIX and Kurucz model coefficients, the light curve residuals exhibited systematic trends near ingress and egress, and the typical reduced chi-squared values increased.  The poor match of theoretical limb darkening coefficients to our data may arise from inaccurate assumptions about the stellar composition, or inaccuracies in the models themselves.  Similar disagreement with model limb darkening has been seen for other high quality data \citep[e.g.][]{knutson07}. 

Since incorrect limb darkening coefficients can introduce systematic bias in the measured transit depths, we chose to estimate the limb darkening coefficients directly from the data.  Our light curves are not precise enough to constrain a two-parameter limb darkening model, so we instead fit for a single linear parameter in each channel.  The data are sufficiently precise to distinguish between the transit depth and the limb darkening coefficient, as evidenced by the lack of correlation between those parameters in the pairs plot shown in Figure\,\ref{fig:pairslk}.

\subsection{Correction of Dilution from Stellar Companions}
\label{subsec:depth_meas}
\begin{figure}
\resizebox{\hsize}{!}{\includegraphics{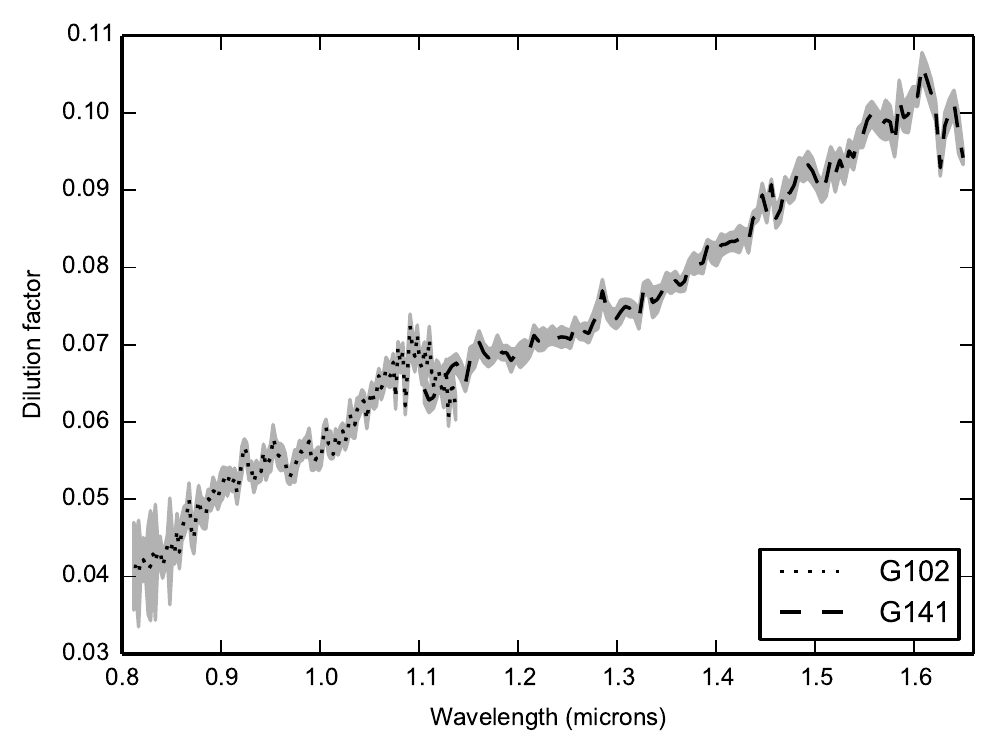}}
\caption{The dilution factor $\alpha_\mathrm{Comp}$, determined from the ratio of flux from WASP-12\,BC to WASP-12\,A.  The median dilution (over the 10 staring mode observations) is shown with black lines (dotted for G102 and dashed for G141).  The gray shaded region indicates 1$\sigma$ uncertainty, determined from the median absolute deviation. To calculate the dilution and uncertainty, we interpolate the spectra from each grism onto a common wavelength scale.}
\label{fig:dilution}
\end{figure}

The transit light curves are affected by dilution from WASP-12\,A's two companion stars, WASP-12\,BC, and from the planet's nightside emission.  Following \cite{stevenson14a}, we calculated a corrected transit depth $\delta^\prime$ for each spectroscopic channel using the formula
\begin{equation}
\delta^\prime = \delta(1 + \alpha_\mathrm{Comp} + \alpha_p)
\end{equation}
where $\delta$ is the measured transit depth, $\alpha_\mathrm{Comp}$ is the ratio of flux from WASP-12\,BC to WASP-12\,A, and $\alpha_p$ is the ratio of the planet's nightside emission to the flux of WASP-12\,A.  We calculated $\alpha_p$ using the same model as \cite{stevenson14a}.  We determined $\alpha_\mathrm{Comp}$ empirically using the spectra extracted from the staring mode observations.  Each visit is assigned a unique $\alpha_\mathrm{Comp}$ value to account for differences in the orientation of the spectra on the detector.  We plot the dilution factor as a function of wavelength in Figure\,\ref{fig:dilution} (calculated by interpolating the companion spectrum to the same wavelength scale as the WASP-12\,A spectrum).  The measured dilution is roughly 10\% larger than the values used by \cite{stevenson14a}; this difference could result from systematic uncertainty in the aperture photometry used to scale the dilution \citep{stevenson14a}.  

\begin{figure*}
\resizebox{1.0 \textwidth}{!}{\includegraphics{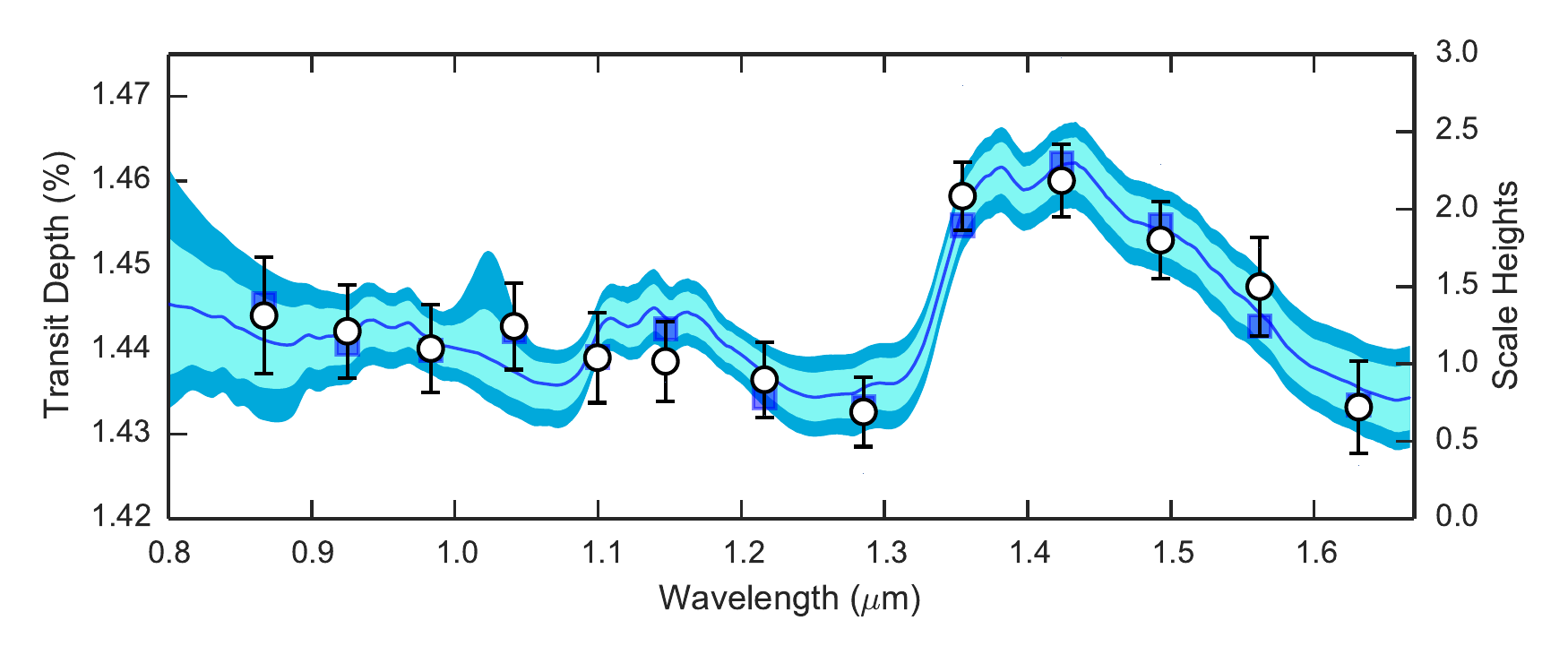}}
\caption{The transmission spectrum of WASP-12b measured with \textit{HST}/WFC3 (points).  The error bars on the transit depths are 1\,$\sigma$ uncertainties from an MCMC fit to the light curves.  We show the best fit model binned to the resolution of the data (blue squares).  The shaded regions indicate 1- and 2\,$\sigma$ credible intervals in the retrieved spectrum (medium and light blue, respectively), relative to the median fit (dark blue line).  The secondary y-axis labels indicate the atmospheric scale height.  One scale height corresponds to 470 km (for a temperature of 1400 K).  The increase in transit depth near 1.4\,$\mu$m corresponds to an H$_2$O bandhead.}
\label{fig:spectrum}
\end{figure*}

To obtain final transit depths, we take the weighted average of the corrected transit depths from each visit.  Uncertainties in the dilution factor corrections are propagated through to the final transit depth uncertainties.  We report the corrected transit depth measurements in Table\,\ref{tab:transit_depths}, and we show the transmission spectrum in Figure\,\ref{fig:spectrum}.  The corrected transit depths are consistent between the visits, indicating that the estimated uncertainties are appropriate, and confirming that stellar activity does not significantly influence the measured spectrum.

\section{Comparison with Previous Results}
\label{sec:comparison}
\begin{figure}
\resizebox{\hsize}{!}{\includegraphics{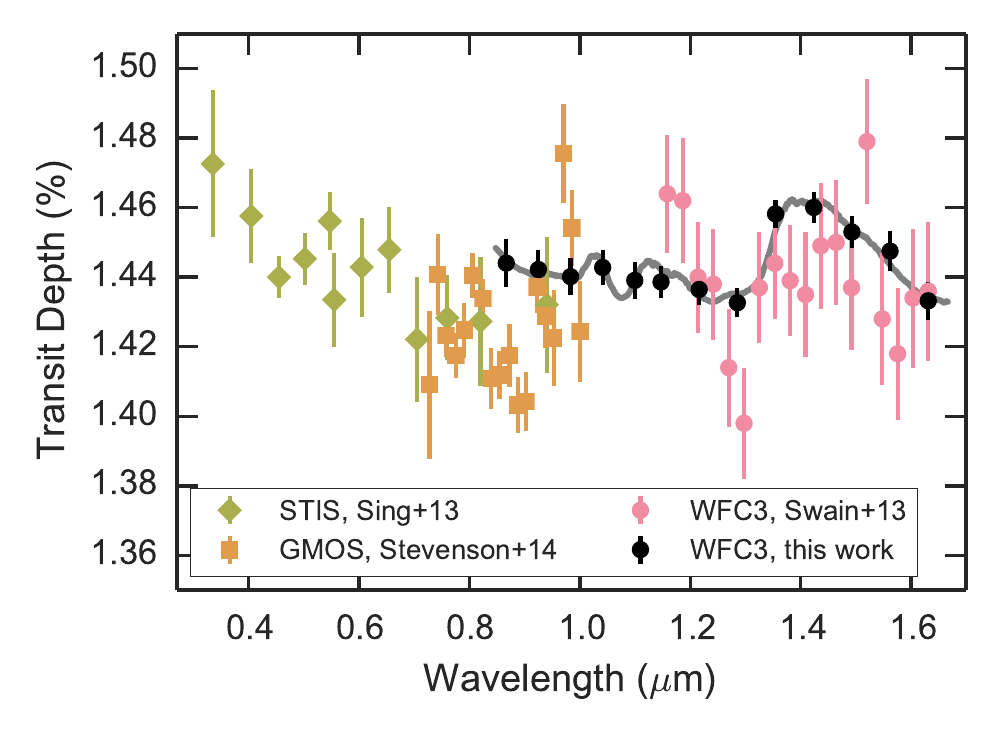}}
\caption{High-precision optical/near-IR transit depth measurements for WASP-12b. We show measurements from \textit{HST}/STIS (green diamonds), \textit{Gemini}/GMOS (orange squares), and \textit{HST}/WFC3 (from this work, black circles; from staring mode data, pink circles).  The best fit model from the CHIMERA FULL retrieval is indicated by the gray line.  Note that this model was fit to the WFC3 data only.  There may be small offsets between data sets due to different corrections for the companion star dilution and the challenge of measuring absolute transit depths in the presence of systematic errors \citep{stevenson14a, stevenson14d}.}
\label{fig:alldata}
\end{figure}

There are several other high-precision transit depth measurements for WASP-12b in addition to the WFC3 spectrum we report here.  The most precise of these are spectroscopy between 0.34 and 0.94 $\mu$m from \textit{HST}/STIS \citep{sing13}, spectroscopy between 0.73 and 1.00 $\mu$m from \textit{Gemini}/GMOS \citep{stevenson14a}, a staring mode \textit{HST}/WFC3 G141 spectrum from \cite{swain13}, and photometry at 3.6 and 4.5 $\mu$m from \textit{Spitzer}/IRAC \citep{cowan12}.  Figure\,\ref{fig:alldata} shows past optical/near-IR transmission spectrum measurements compared to our results.

Our transit depth measurements are consistent with the staring mode spectrum from \textit{HST}/WFC3 \citep{swain13, mandell13, stevenson14a}, modulo a constant offset in the absolute transit depths.  We chose not to incorporate this data set in our analysis, however.  The staring mode data has a low duty cycle and increases the amount of in-transit exposure time by just 10\% over the total from our three spatial scan observations.  Morever, there is uncertainty about the best instrument systematics model for this data set \citep{stevenson14a}.  Since the marginal improvement in measurement precision is counteracted by increased systematic uncertainty, we chose to focus on the transmission spectrum derived from the spatial scan data only. 

Our spectrum agrees less well with the ground-based transmission spectrum from \textit{Gemini}/GMOS \citep{stevenson14a}.  To compare the WFC3 and GMOS results, we computed $\chi^2$ values for the data using the best fit model spectrum from the FREE retrieval (described in \S\,\ref{subsec:FREE}) over the wavelength range where the spectra overlap (0.86 - 1.00 $\mu$m).  We allowed a free offset in the model for the GMOS data to account for differences in the absolute measured transit depths.  We find that the GMOS data are inconsistent with the model at $> 4\,\sigma$ ($\chi^2 = 38.7$ for 9 degrees of freedom).  Based on the track record of precise, reproducible transmission spectra from WFC3 that are well-fit by theoretical models \citep[e.g.][]{kreidberg14a, kreidberg14b, stevenson14c}, we trust the reliability of the WFC3 measurements over those from GMOS.  Ground-based transit observations exhibit strong time-correlated noise that makes estimating accurate confidence intervals challenging.  Additional GMOS transit spectroscopy observations would help identify the number of repeated observations with this instrument needed to make robust measurements. 

The \textit{Spitzer} transit depths measurements are approximately $0.1\%$ smaller than the mean WFC3 transit depth.  Based on the random errors alone, this difference is significant.  However, the \textit{Spitzer} data have significant systematic uncertainties, because the relative transit depths measured with IRAC channels 1 and 2 can shift by $>0.1$\% depending on the aperture size used in the data reduction. There is no obvious optimal choice of aperture \citep[see Figure 17,][]{stevenson14a}.  We therefore chose not to incorporate the Spitzer results in our analysis.

\begin{deluxetable}{cccc}
\tabletypesize{\scriptsize}
\tablecolumns{4}
\tablewidth{0pc}
\tablecaption{Transit Fit Parameters}
\tablehead{ \colhead{Bandpass\tablenotemark{a}} & \colhead{(R$_{p}$/R$_{\star}$)$^{2}$\tablenotemark{b}} & \colhead{Limb-} & \colhead{rms} \\
($\mu$m) & (\%) &darkening $u$ & \colhead{(ppm)}  }
\startdata
0.838 -- 0.896 & 1.4441 $\pm$ 0.0069 & 0.27 $\pm$ 0.02 & 349 \\
0.896 -- 0.954 & 1.4422 $\pm$ 0.0055 & 0.28 $\pm$ 0.01 & 300 \\
0.954 -- 1.012 & 1.4402 $\pm$ 0.0052 & 0.27 $\pm$ 0.01 & 296 \\
1.012 -- 1.070 & 1.4428 $\pm$ 0.0051 & 0.26 $\pm$ 0.01 & 301 \\
1.070 -- 1.129 & 1.4391 $\pm$ 0.0053 & 0.25 $\pm$ 0.01 & 299 \\[0.8mm]
\hline\\
1.112 -- 1.182 & 1.4386 $\pm$ 0.0047 & 0.26 $\pm$ 0.01 & 267 \\
1.182 -- 1.251 & 1.4365 $\pm$ 0.0045 & 0.26 $\pm$ 0.01 & 267 \\
1.251 -- 1.320 & 1.4327 $\pm$ 0.0041 & 0.21 $\pm$ 0.01 & 253 \\
1.320 -- 1.389 & 1.4582 $\pm$ 0.0040 & 0.22 $\pm$ 0.01 & 242 \\
1.389 -- 1.458 & 1.4600 $\pm$ 0.0043 & 0.18 $\pm$ 0.01 & 255 \\
1.458 -- 1.527 & 1.4530 $\pm$ 0.0045 & 0.20 $\pm$ 0.01 & 270 \\
1.527 -- 1.597 & 1.4475 $\pm$ 0.0058 & 0.16 $\pm$ 0.02 & 332 \\
1.597 -- 1.666 & 1.4332 $\pm$ 0.0055 & 0.16 $\pm$ 0.02 & 321 
\enddata
\tablenotetext{a}{The measurements between 0.838 - 1.129 $\mu$m are from the G102 grism; those from 1.112 - 1.666 $\mu$m are from the G141 grism.}
\tablenotetext{b}{Transit depths corrected for dilution from companion stars and planet nightside emission.  The light curve fits had $a/R_s$ and $i$ fixed to 2.91 and 80.56$^\circ$, respectively. }
\label{tab:transit_depths}
\end{deluxetable}		

\section{Retrieval of the Atmospheric Properties}
\label{sec:retrieval}
Given the high precision of the WFC3 transmission spectrum, we can put more powerful constraints on the planet's atmosphere than has been possible with past measurements.  No previous data set for WASP-12b has shown conclusive spectroscopic evidence for molecular absorption, but the WFC3 spectrum has a noticeable increase in transit depth near the center of the water absorption band at 1.4 $\mu$m (see Figure\,\ref{fig:spectrum}).

To quantify the water abundance and other atmospheric properties based on the transmission spectrum, we performed a retrieval using the CHIMERA suite \citep{line13a,  line14}. Modifications to the retrieval code and description of the transmission forward model are provided in \cite{line13b, stevenson14c, kreidberg14b, swain14, diamond-lowe14}.  We tested two different model parameterizations with CHIMERA.  The first is the widely adopted approach of retrieving molecular abundances without any constraints from chemistry \citep[e.g.][]{madhusudhan09}.  For our second approach, we developed a new parameterization that retrieves C/O and atmospheric metallicity under the assumption of chemical equilibrium.  We refer to these methods as the FREE approach and the CHEMICALLY-CONSISTENT (C-C) approach, respectively.  Both methods are described in more detail below. We also present a comparison with results from the NEMESIS retrieval code \citep{irwin08}.

We elected to focus on the WFC3 data for our retrievals, based on the caveats for other data described in \S\,\ref{sec:comparison}.  We treated the spectra from the two WFC3 grisms as a single data set, with no offset in transit depth between them.  The transit depths are consistent between the grisms because the influence of stellar activity is below the measurement precision (see \S\,\ref{subsec:observations}), so no offset is needed.  We also tested including STIS data in some of the retrievals (see \S\,\ref{subsubsec:otherstuff}), but found that it does not affect the main conclusions of this work.

\subsection{FREE Retrieval with CHIMERA}
\label{subsec:FREE}
The FREE parameterization retrieves molecular abundances, cloud and haze properties, and the altitude-independent scale height temperature.

Our model explored the dominant molecular opacities expected for a solar composition gas at the temperatures and pressures probed by the observations.  These are H$_2$O, TiO, Na, and K over the wavelength range of the WFC3 spectrum \citep{burrows99, fortney08}.  The remaining gas is assumed to be a solar composition mixture of H$_2$ and He. We discuss results for including additional molecular species in \S\,\ref{subsubsec:otherstuff}.

In addition to the molecular opacities, we included opacity from clouds and haze.  Previous analyses of WASP-12b's transmission spectrum have suggested that hazes are present in the atmosphere, but the haze composition is poorly constrained \citep{sing13,stevenson14a}.  We therefore used a flexible model that includes a power law haze and an opaque gray cloud deck.  We modeled clouds as a gray opacity source that masks transmission through the atmosphere below a fixed pressure level $P_c$.  The haze opacity was parameterized by $\sigma(\lambda) = \sigma_0 (\lambda/\lambda_0)^\gamma$, where the scattering amplitude $\sigma_0$ and slope $\gamma$ are free parameters \citep[as in][]{lecavelier08}.  The scattering is presumed to happen throughout the entire atmosphere as if it were a well mixed gas. The scattering amplitude is a scaling to the H$_2$ Rayleigh scattering cross section times the solar H$_2$ abundance at 0.43 $\mu$m.  More sophisticated models for clouds and haze would require additional free parameters that are not justified by the precision of our data.  Our simple parameterization is sufficient to capture degeneracies between clouds/haze and the water abundance, which is the primary goal of this investigation.

We also retrieved a ``scale height temperature" parameter $T_s$.  We assumed the atmosphere is isothermal, with temperature equal to $T_s$ at all pressures. We tested fitting for a more complex temperature-pressure profile but found it did not affect our results because the transmission spectrum is only weakly sensitive to the atmosphere's thermal structure \citep[see][]{barstow13, barstow14}.

The final free parameter in the retrieval was a scale factor for the planet's 10 bar radius.  We assumed a baseline planet radius of 1.79 $R_\mathrm{J}$ and a stellar radius of 1.57 $R_\odot$ from \cite{hebb09}.  The planet radius was scaled by a factor $R_\mathrm{scale}$ to account for uncertainty in the pressure level in the atmosphere at a given radius.  To first order, the effect of this scaling is to shift the model transit depths by a constant factor.  A second order effect is that scaling the radius changes the amplitude of spectral features \citep[see Equation 1 in][]{lecavelier08}.

\begin{figure}
\resizebox{\hsize}{!}{\includegraphics{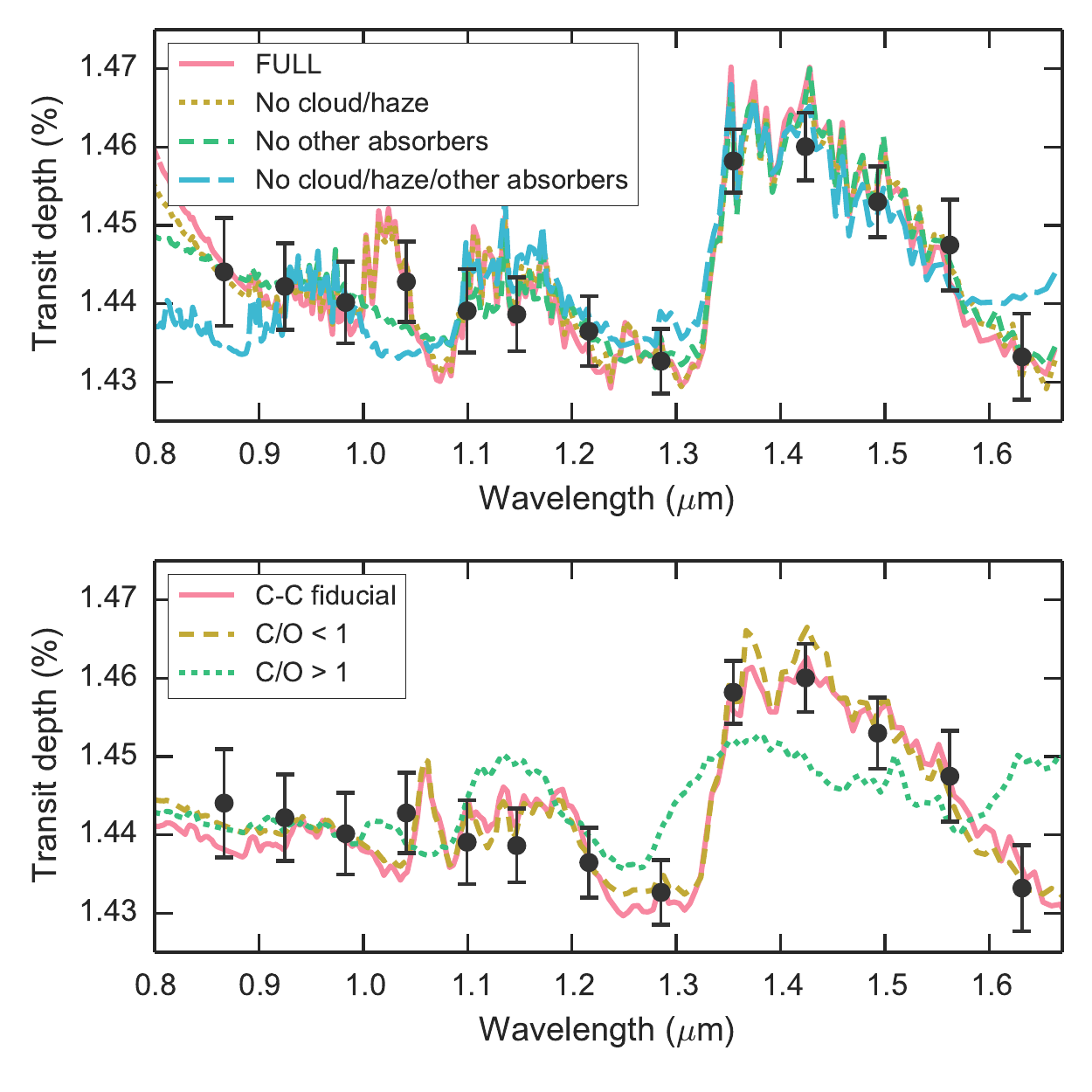}}
\caption{Best fits for different assumed models (lines) compared to the WFC3 transmission spectrum (points).  Models are binned to a wavelength resolution of 0.01 $\mu$m.  Results from the FREE and C-C parameterizations are shown in the top and bottom panels, respectively.}
\label{fig:bestfits}
\end{figure}

\begin{figure*}
\begin{centering}
\resizebox{1.0\textwidth}{!}{\includegraphics{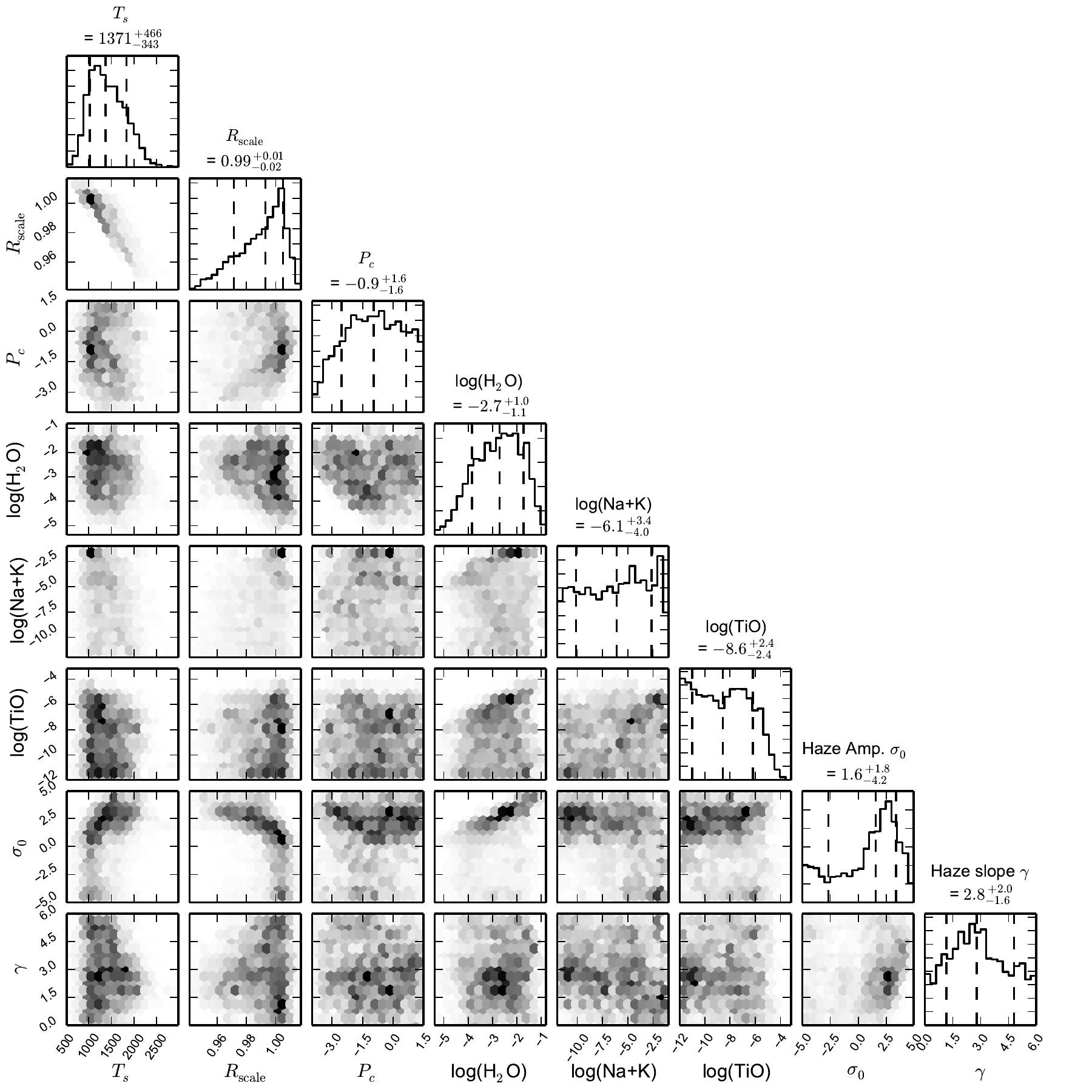}}
\caption{A pairs plot of the distribution of parameters retrieved with CHIMERA for the FULL model.  The parameters are the scale height temperature $T_s$ (in Kelvin), the planet radius scale factor $R_\mathrm{scale}$, the cloud-top pressure $P_c$ (in log bars), the logarithm (base 10) of the molecular abundances, and the haze scattering amplitude $\sigma_0$ and slope $\gamma$. The off-diagonal plots show marginalized posterior probability density for pairs of parameters (darker shading corresponds to higher probability).  The diagonal plots show marginalized posterior probability distributions for individual parameters, with the median and 68\% credible interval marked with dashed lines.}
\label{fig:pairs_ml}
\end{centering}
\end{figure*}

In sum, the retrieval had 8 free parameters:  the abundances of H$_2$O, Na+K (fixed at the solar abundance ratio), and TiO, as well as clouds and haze, the scale height temperature, and the planet radius scale factor. We refer to this 8-parameter fit as the FULL model.  The best fits for the FULL model and nested models within it are shown in Figure\,\ref{fig:bestfits}. The best-fit models are those that produced the lowest $\chi^2$ values in the MCMC. We list the retrieved water abundances, temperatures, and best fit $\chi^2$ values for these models in Table\,\ref{tab:significances}.  The distributions of retrieved parameters from the FULL model are shown in Figure\,\ref{fig:pairs_ml}.  We note that all of the nested models are nearly indistinguishable near the water absorption feature at 1.4 $\mu$m. The best constrained molecular abundance is that of water: we retrieved a water volume mixing ratio (VMR) of $1.6\times10^{-4} - 2.0\times10^{-2}$ at 1\,$\sigma$, which is consistent with expectations for a solar composition gas.  The other molecular abundances are not as well constrained.  We retrieved a 3\,$\sigma$ upper limit on the VMR of TiO equal to $2\times10^{-4}$, and the abundance of Na+K is unbounded.  The cloud and haze properties are also poorly constrained.  We discuss the implications of these measurements for the atmosphere composition in detail in \S\,\ref{sec:ctoo}.  

\subsubsection{Detection Significances}
We performed nested model selection to identify how strongly each opacity source is detected in the spectrum.  The standard Bayesian approach for comparing models is to use Bayes factors, which can be converted to detection significances \citep{sellke01, trotta08}.  See also \cite{benneke13} for an application of nested Bayesian model comparisons as applied to super-Earth atmospheres.  Bayes factors are the ratio of the Bayesian evidence (marginal likelihood) of the two models under consideration.  The evidence is a multidimensional integral over the entire posterior volume, a non-trivial calculation, so here we use two methods to approximate the integral. The first is the Numerical Lebesgue Algorithm described in \cite{weinberg12} \citep[see also][for an application to exoplanet atmospheres and a comparison to the $\Delta\chi^2$ test]{swain14}. The second is the Laplace approximation \citep{kass95, cornish07}. Most methods for computing Bayesian evidence diverge for low detection significances, but they agree well for highly significant detections \citep[see Fig. 3 from][]{cornish07}.

\begin{deluxetable*}{lcr@{}c@{}lccc}
\tabletypesize{\scriptsize}
\tablecolumns{6}
\tablewidth{0pc}
\tablecaption{FREE Retrieval Results}
\tablehead{ 
\colhead{Scenario\tablenotemark{a}} & \colhead{Water abundance\tablenotemark{b}} & \multicolumn{3}{c}{Temperature\tablenotemark{b}} & \colhead{$\chi^2$\tablenotemark{c}} & \colhead{Weinberg} & \colhead{Laplace} \\
  & \colhead{(VMR)} & \multicolumn{3}{c}{(Kelvin)} & & \colhead{significance\tablenotemark{d} ($\sigma$)} & \colhead{significance ($\sigma$)}} \\
\startdata
FULL ($T_s$, $R_\mathrm{scale}$, $P_c$, H$_2$O, TiO, Na+K, $\sigma_0$, $\gamma$) & $1.5\times10^{-4} - 2.2\times10^{-2}$ & 1040 &$\,-\,$& 1870 & 2.82 & -- & -- \\
No water ($T_s$, $R_\mathrm{scale}$, $P_c$, TiO, Na+K, $\sigma_0$, $\gamma$) & -- & 130 &$\,-\,$ & 1560 & 47.0 & 6.9 & 7.1 \\
No cloud, haze, other absorbers ($T_s$, $R_\mathrm{scale}$, H$_2$O) & $1.1\times10^{-5}-1.5\times10^{-2}$ & 730 &$\,-\,$& 1170 & 11.1 & 3.2 & 4.6 \\
No other absorbers ($T_s$, $R_\mathrm{scale}$, $P_c$, H$_2$O, $\sigma_0$, $\gamma$) & $6.9\times10^{-5}-2.9\times10^{-2}$ & 1090 &$\,-\,$& 1890 &3.59 & 1.2 & 2.6 \\
No haze ($T_s$, $R_\mathrm{scale}$, $P_c$, H$_2$O, TiO, Na+K) & $5.8\times10^{-5}-8.3\times10^{-3}$ & 910 &$\,-\,$& 1470 &2.87 & 1.6 & 2.5 \\
No cloud ($T_s$, $R_\mathrm{scale}$, H$_2$O, TiO, Na+K, $\sigma_0$, $\gamma$) & $6.6\times10^{-5}-2.0\times10^{-4}$ & 1000 &$\,-\,$& 1840 & 2.86 & -- & 2.0 \\
No cloud/haze ($T_s$, $R_\mathrm{scale}$, H$_2$O, TiO, Na+K) & $2.4\times10^{-5}-5.5\times10^{-3}$ & 860 &$\,-\,$& 1430 &2.95 & -- & 1.2 
\enddata
\tablenotetext{a}{Indicates the parameters that are removed from the FULL model.  The remaining parameters are shown in parentheses.}
\tablenotetext{b}{The range of values corresponds to the 68\% credible interval centered on the median retrieved value.}
\tablenotetext{c}{$\chi^2$ values are calculated for the best fit model.}
\tablenotetext{d}{Scenarios where the detection significance is undefined (because the Bayes factor is less than 1) are marked by --. The detection significance for the FULL model is also undefined because significances are defined relative to it.}
\label{tab:significances}
\end{deluxetable*}		

Using these techniques, we determined the detection significances for the following parameter combinations: H$_2$O, absorbers other than H$_2$O, clouds and haze combined, clouds only, haze only, and all opacity sources other than H$_2$O. We computed the detection significance for each nested model by calculating its Bayes factor relative to the FULL model. The results are shown in Table\,\ref{tab:significances}.  If a parameter has a low detection significance value ($\lessapprox3\,\sigma$), that parameter does not provide a statistically significant improvement to the model fit.  

We find that H$_2$O is detected at high confidence in the spectrum ($7\,\sigma$).  We also find that the presence of all other opacity sources combined (clouds, haze, TiO, Na and K) is significant at $>3\,\sigma$.  However, no other combination of these opacities is significantly detected (e.g., the detection significance for clouds alone is $<2\,\sigma$).  The reason for this is that the parameters are degenerate.  For example, both clouds and haze both have the effect of truncating the height of the water feature.  Similiarly, haze, the wings of the alkali metal lines, and presence of TiO can all contribute to the rise in transit depths towards the blue end of the spectrum. Therefore, if a few of these opacity sources are removed, the others can compensate. However, removing all opacities besides water results in a significantly poorer model fit (as can be seen in Figure\,\ref{fig:bestfits}).  We therefore focus on results from the FULL model because it accounts for the likely presence of some combination of these other species and the degeneracies they may have with H$_2$O. 

\subsubsection{Retrievals with Additional Molecules/Data} 
\label{subsubsec:otherstuff}
We explored many different combinations of models and data sets before we arrived at the analysis presented above.  One option we considered was to fit a WFC3 spectrum with smaller wavelength bins.  We created a ``high-resolution" spectrum with $R = 55 - 70$, in contrast to the ``low-resolution" data ($R = 15 - 25$) we used for our final modeling.  The spectra are qualitatively similar, but the models fit the low-resolution data better ($\chi^2_\nu$ = 0.6 versus 1.5).  The high-resolution spectrum has larger random scatter around the best fit model, but no systematic trends indicating that the model is missing a particular physical effect.  The scatter could be due to undiagnosed systematics at the pixel level in the data, which is remedied by increasing the number of pixels that contribute to each spectral channel.  

The retrieved water abundances for both resolutions are very similar ($1.5\times10^{-4}-2.2\times10^{-2}$ and $1.2\times10^{-4} - 1.8\times10^{-2}$ for low- and high-resolution).  On the other hand, the best fit scale height temperatures differ by 1.4\,$\sigma$ ($1370^{+470}_{-340}$  versus $2020^{+320}_{-340}$ for low/high). This difference is due to slight changes in the morphology of the water feature between the data sets. It illustrates how retrieval results can be biased or have underestimated uncertainties when the model does not provide a good fit to the data, either because of missing model physics or underestimated data error bars.

Another test we performed was to fit the combined transmission spectrum from \textit{HST}/STIS and WFC3.  The STIS spectrum covers optical wavelengths and exhibits a linear increase in transit depth from red to blue (see Figure\,\ref{fig:alldata}).  Incorporating the STIS data in the retrieval increased the detection significance of clouds and haze to $>3\,\sigma$ (because of the rise in transit depth towards the blue), but otherwise did not significantly change the results.

We also tested including additional absorbing species in the retrieval.  Previous work has suggested WASP-12b has a carbon-rich atmosphere composition \citep[e.g.][]{madhusudhan11a}, so we wanted to confirm that our estimate of the water abundance is not biased by only including absorbers expected for solar composition.  We therefore ran a retrieval including all the major opacity sources expected for either oxygen-rich or carbon-rich compositions, including H$_2$O, CO, CO$_2$, NH$_3$, TiO, VO, Na, K, CH$_4$, C$_2$H$_2$, HCN, H$_2$S, FeH, N$_2$, and collisionally-induced H$_2$/He absorption \citep{burrows99, fortney08, madhusudhan12}.  For this retrieval, we also fit the STIS data and used the higher-resolution WFC3 spectrum.  The retrieved water abundance of $2.5\times10^{-4}-2.0\times10^{-2}$ is nearly identical to the results we obtained from the FULL model.  No individual absorbers were significantly detected besides H$_2$O.

Our conclusion from all the scenarios we fit is that the constraints on the water abundance are not significantly impacted by the choice of data sets or the absorbers included in the modeling.

\subsection{CHEMICALLY-CONSISTENT Retrieval with CHIMERA}
In addition to retrievals with the FREE parameterization (described in \S\,\ref{subsec:FREE}), we also developed a reparameterized model to retrieve atmospheric properties that are consistent with chemical equilibrium.  Rather than varying the mixing ratios of individual gases, we made C/O and metallicity free parameters and computed thermochemical equilibrium abundances along the temperature/pressure (T/P) profile for major species.  We retrieved the T/P profile rather than a scale height temperature in order to explore any additional degeneracies due to the profile shape.  The T/P profile is modeled with a 5-parameter analytic function \citep{line13a} which produces physically realistic thermal profiles consistent with radiative equilibrium \citep[e.g.][]{guillot10, robinson12, heng12, parmentier14}.  We calculated chemical profiles for H$_2$, He, H$_2$O, CH$_4$, CO, CO$_2$, NH$_3$, H$_2$S, PH$_3$, C$_2$H$_2$, HCN, Na, K, FeH, TiO, VO, and N$_2$ to cover the full range of gases that could contribute significant opacity in the near-IR. The equilibrium abundances are computed using the NASA Chemical Equilibrium with Applications code \citep{mcbride96, moses11, line11}. The chemical profiles were fed into the radiative transfer model to calculate model transmission spectra for comparison with our data.  The model also included opacity from clouds and haze (using the same formalism described in \S\,\ref{subsec:FREE}), and the reference radius $R_p$, for a total of 11 free parameters.  While we reduced the number of free absorber parameters to just C/O and metallicity, the overall number of free parameters increased due to the more flexible T/P profile.

\begin{deluxetable*}{lr@{}c@{}lccr@{}c@{}lc}
\tabletypesize{\scriptsize}
\tablecolumns{7}
\tablewidth{0pc}
\tablecaption{C-C Retrieval Results}
\tablehead{\colhead{Scenario} & \multicolumn{3}{c}{C/O} & \colhead{Metallicity} & \colhead{Water abundance} & \multicolumn{3}{c}{Temperature\tablenotemark{a} (K)} & \colhead{$\chi^2$}\\
  & \multicolumn{3}{c}{}  & \colhead{($\times$\,solar)} & \colhead{(1\,$\sigma$)} & \multicolumn{3}{c}{(1\,$\sigma$)} & }\\
\startdata
Fiducial  & $0.2$ & $\,-\,$ & $0.7$ & $0.3-20$ & $9.3\times10^{-5}$ - $5.3\times10^{-3}$ & 1090 & $\,-\,$ & 1760 & 2.43\\
O-rich    & $0.1$ & $\,-\,$ & $0.7$ & $0.3-30$ & $1.2\times10^{-4}$ - $1.5\times10^{-2}$ & 1090 & $\,-\,$ & 1890 & 2.75 \\
C-rich    & $2.9$ & $\,-\,$ & $51 $ & $0.2-80$ & $1.5\times10^{-5}$ - $8.6\times10^{-3}$ & 120  & $\,-\,$ & 550 & 29.2
\enddata
\tablenotetext{a}{The temperature range corresponds to a pressure level of 1 mbar.}
\label{tab:selfcons}
\end{deluxetable*}		
 
\begin{figure*}
\resizebox{\hsize}{!}{\includegraphics{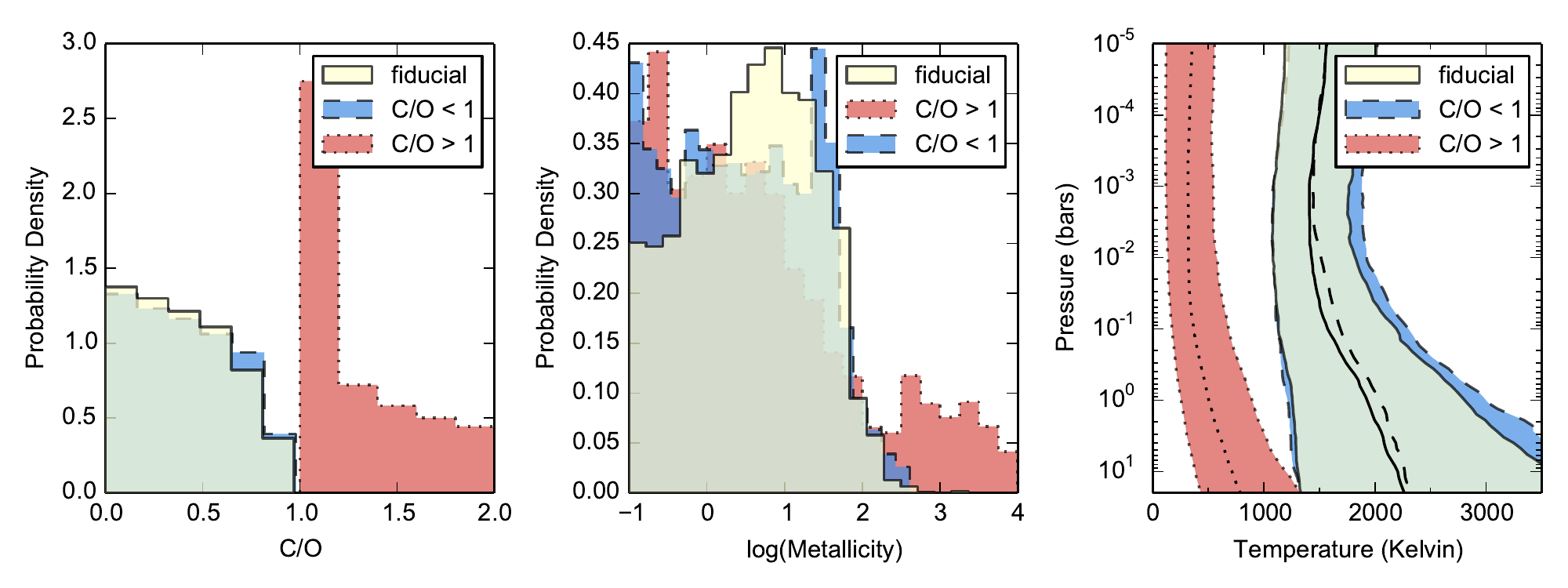}}
\caption{Retrieval results from the C-C parameterization. The left and middle panels show the marginalized distribution for C/O and metallicity.  The right panel shows the median (black lines) and 1\,$\sigma$ range (shaded regions) of 1000 randomly sampled temperature-pressure profiles for each scenario.  Yellow, blue, and red shading correspond to constraints from the fiducial scenario (an uninformative prior on C/O), the oxygen-rich scenario (C/O\,$<1$), and a carbon-rich scenario (C/O\,$>1$).  The distribution of C/O values for the carbon-rich model is normalized to have a probability mass of unity over the plotted range; however, the distribution has an extended tail toward higher C/O values that is not shown.}
\label{fig:selfcons}
\end{figure*}

We ran retrievals using three different priors for C/O.  The fiducial case had an uninformative prior constraint on C/O.  We also ran retrievals that constrained the atmospheric composition to be either oxygen-rich or carbon-rich.  For the O-rich scenario, the prior probability was set to zero for C/O values greater than unity.  Correspondingly, the C-rich scenario had zero prior probability for C/O\,$<1$.  We show the best fit models for all three cases in Figure\,\ref{fig:bestfits}. Figure\,\ref{fig:selfcons} shows marginalized distributions of C/O and metallicity for each scenario, as well as constraints on the temperature-pressure profiles. Table\,\ref{tab:selfcons} gives the $\chi^2$ values of the best fits in each scenario.  It also lists the 68\% credible intervals for C/O, metallicity, and the temperature and water abundance at 1 mbar pressure. 

The fiducial and oxygen-rich scenarios give nearly identical constraints on C/O and the atmospheric metallicity.  Even though the fiducial model has an uninformative prior on C/O, 100\% of the retrieved C/O values are less than unity for this case (suggesting the result is data-driven rather than prior-driven).  Both scenarios yield constraints on the molecular abundances and thermal profile that agree well with results from the FULL model.  The median retrieved water abundances are $6.3\times10^{-4}$ and $1.0\times10^{-3}$ for the fiducial and O-rich scenarios, respectively.  The median 1 mbar temperatures are 1410 and 1450 K.  We retrieved metallicities in the range $0.3 -30\times$\,solar (at 1\,$\sigma$). The retrieved cloud and haze properties for these scenarios are consistent with the FULL model at 1\,$\sigma$. 

\begin{figure*}
\resizebox{1.0\textwidth}{!}{\includegraphics{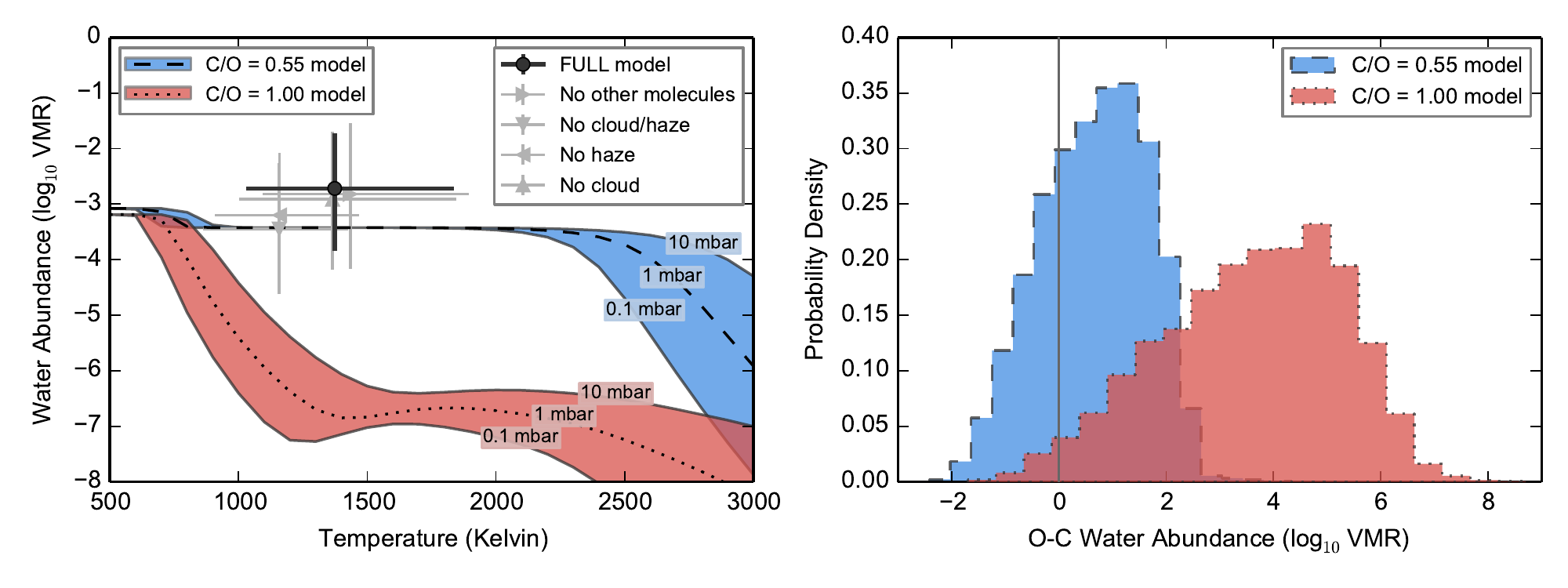}}
\caption{\textbf{Left:} Measurement of the water abundance and scale height temperature for WASP-12b (points) compared to equilibrium chemistry predictions of the water content for different atmospheric compositions (lines and shading).  The black point indicates the water abundance and temperature measurements from the FULL model from the CHIMERA fit to the WFC3 spectrum.  Results from other nested models are shown in gray.  The black dashed line and blue shading correspond to water abundance predictions for a solar C/O composition, and the black dotted line and red shading correspond to C/O = 1.  Both models have solar metallicity.  For each model composition, the shading shows the span of predicted water abundances over pressures ranging from 0.1 - 10 mbar. The black lines correspond to 1 mbar, which is the typical pressure level probed by our observations. \textbf{Right:} Histogram of observed minus calculated water abundances for the FULL retrieval results relative to different model compositions.  The red histogram (dotted line) shows the comparison with a carbon-rich model, and the blue histogram (dashed line) shows results for the oxygen-rich model.  The gray vertical line marks zero, where the observations are an exact match to the model predictions.}
\label{fig:waterpred}
\end{figure*}

By contrast, the carbon-rich scenario produced significantly different atmospheric properties.  To reproduce the water absorption feature in the spectrum, the temperature was driven to much lower values (the median is 320 K at 1 mbar).  These low temperatures allow for higher water abundances (see Figure\,\ref{fig:waterpred}), though they are unlikely for the terminator region given how highly irradiated the planet is.  The median water abundance, $1.4\times10^{-4}$, is comparable to that for the O-rich and fiducial scenarios. On the other hand, the median methane abundance increases to $2.4\times10^{-3}$, versus $1.3\times10^{-11}$ for the fiducial case.  These differences have two main effects on the model transmission spectra.  One is that absorption features have smaller amplitude because lower temperatures decrease the atmospheric scale height.  The second effect is that methane absorption is present in the spectrum.  It is especially noticeable in the window between water features at $>1.6\,\mu$m wavelengths (see the best-fit C-rich model in Figure\,\ref{fig:bestfits}).  The consequence of these changes is that C-rich models do not fit the measured spectrum as well as O-rich models.  We discuss the strength of the evidence for one scenario over the other in quantitative detail in \S\,\ref{sec:ctoo}. 

As a test, we also considered a scenario with clouds and haze removed. We computed the Bayes factor for this nested model and found a detection significance of 3.7\,$\sigma$ for the cloud and haze parameters.  The detection significance is higher than for the FREE model parameterization because the assumption of chemical equilibrium breaks the degeneracy between the cloud/haze and the other molecular abundances (Na+K, TiO); i.e. the other molecular abundances can't increase arbitrarily to make up for the absence of clouds/haze.

\subsection{Retrieval with NEMESIS}
We also compared the results from CHIMERA with output from an independent retrievel code to test the robustness of our measurements against a different modeling approach (optimal estimation versus MCMC).  We fit the high-resolution WFC3 spectrum with the NEMESIS code \citep{irwin08}.  The model included H$_2$O as the only molecular opacity source.  We modeled the atmosphere's thermal structure as a scalar multiple of the dayside temperature-pressure profile from \cite{stevenson14b}.  NEMESIS does not currently incorporate cloud-top pressure as a free parameter, so we fixed the altitude of an opaque gray cloud deck over a grid of pressures ranging from 10 to $10^{-3}$ mbar and retrieved the atmospheric properties for each case \citep[for further discussion of retrieving cloud properties with NEMESIS, see][]{barstow13}.  The best fit model had a cloud deck at 1 mbar and a retrieved water abundance of $4.0\times10^{-4} - 1.7\times10^{-3}$ at 1\,$\sigma$.  Models with higher altitude cloud decks (0.1 - 0.001 mbar) achieved nearly as good a fit, and had 1 $\sigma$ upper bounds on the water abundance of $3\times10^{-2}$. These results are in excellent agreement with the results from CHIMERA.

\section{Implications for Atmospheric C/O}
\label{sec:ctoo}
In this section, we quantify the strength of the evidence for oxygen-rich compositions (C/O\,$<1$) over carbon-rich compositions (C/O\,$>1$) based on our retrieval results from two separate modeling approaches.  We focus first on constraints from the water abundance from the FREE retrieval, since water is the only molecule that is unambiguously detected in the spectrum.  We then discuss constraints from the C-C model, which retrieves C/O directly under the assumption of chemical equilibrium. 

\subsection{Constraints from the FREE Retrieval}
\label{subsec:FREE_constraints}
Broadly speaking, a carbon-rich atmosphere is expected to have lower water abundance than an oxygen-rich atmosphere, because most of the oxygen atoms are bound in CO in chemical equilibrium at high temperatures \citep{madhusudhan11b}.  Our water abundance measurement is a qualitatively better match to predictions for an oxygen-rich composition.  In Figure\,\ref{fig:waterpred}, we show the retrieved H$_2$O abundance for the FULL retrieval and nested models in comparison with thermochemical equilibrium predictions for oxygen-rich (C/O = 0.55 = solar) and carbon-rich (C/O = 1) atmospheres.  The predicted abundance depends on pressure and temperature, so we plot the span of predictions over pressures from 0.1 to 10 mbar as a function of temperature.

\begin{figure}
\resizebox{\hsize}{!}{\includegraphics{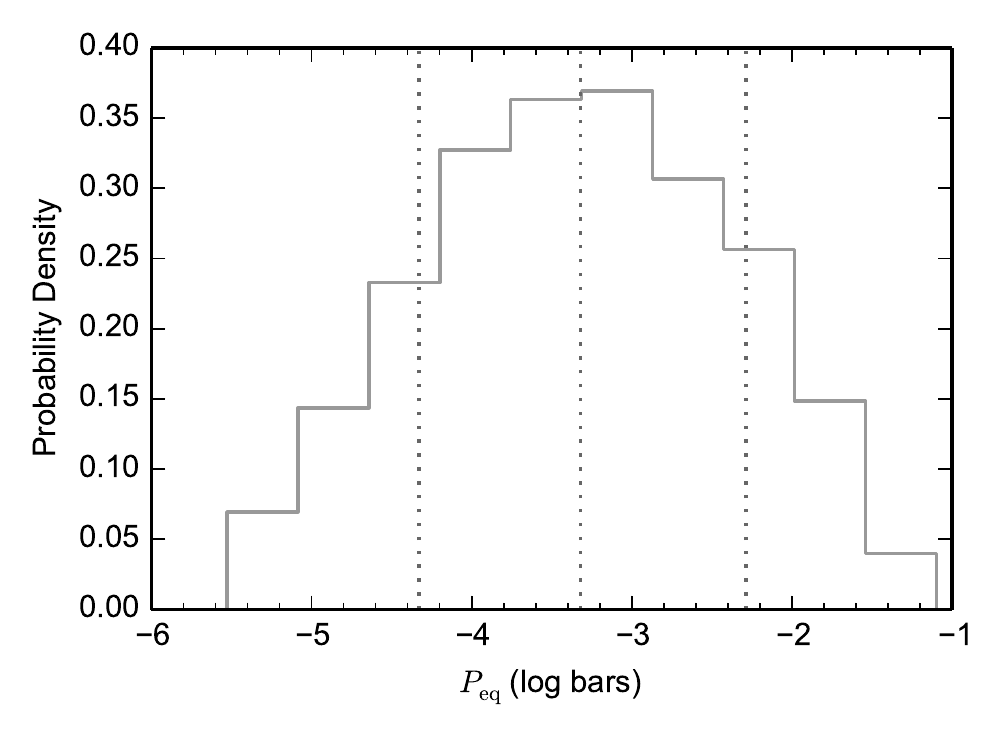}}
\caption{Histogram of equivalent pressures corresponding to an optical depth of 0.56 at 1.4 $\mu$m for each step in the MCMC chain for the FULL retrieval.  The dotted lines indicate the median and surrounding 68\% confidence interval.}
\label{fig:pressures}
\end{figure}

To quantitatively compare our measurement to the models, we must marginalize over the uncertainty in the pressure and temperature probed by the observations.  To do this, we compute an equivalent pressure level $p_{eq}$ corresponding to an optical depth $\tau = 0.56$ at 1.4 $\mu$m at each step in the MCMC chain.  This quantity is representative of the typical pressure level at which photons are absorbed by water molecules \citep{lecavelier08}.  Figure\,\ref{fig:pressures} shows the distribution of equivalent pressures from the MCMC chain.  Note that $p_{eq}$ is not a retrieved quantity, but rather derived from the opacities at each MCMC step.  We show the distribution of equivalent pressures obtained from this method in Figure\,\ref{fig:pressures}.  The pressures have a 1\,$\sigma$ range of 0.05 to 5 mbar, with a peak at 0.5 mbar.

For a specified atmospheric composition, we can then calculate the predicted equilibrium H$_2$O abundance for a temperature and pressure equal to $T_s$ and $p_{eq}$.  This procedure yields a ``calculated" water abundance to compare with the ``observed" water abundance at each step in the MCMC chain.  We show the distribution of observed minus calculated (``O - C") values for the oxygen-rich and carbon-rich models in Figure\,\ref{fig:waterpred} (right panel).  

To test how well each model agrees with the retrieved H$_2$O abundance, we assess whether the O-C distribution is consistent with zero.  For the oxygen-rich model, zero is contained in the 1\,$\sigma$ credible interval centered on the median, indicating that this model is a good match. By contrast, the O-C distribution for the carbon-rich model is in tension with zero at approximately 2$\sigma$ (the 95\% credible interval is $0.1-6.4$). The median O-C value is 3.7, implying that the typical retrieved water abundances are nearly four orders of magnitude larger than predicted for a carbon-rich composition.  

We emphasize that these results are strongly dependent on the temperature and pressure probed by the observations. Assuming a different local thermal profile can result in order-of-magnitude differences in the predicted abundances \citep[cf.\ Figure 2,][]{madhusudhan12}. It is therefore essential to account for the temperature and pressure (and their uncertainties) when estimating C/O.

\subsection{Constraints from the C-C Retrieval}
We obtain more stringent constraints on the atmospheric C/O from the C-C retrieval than from the FREE parameterization.  By every metric we enumerate below, C/O values greater than one are ruled out at high confidence.  First, the entire posterior probability distribution for C/O from the fiducial model is less than one.  We calculate that there are roughly 3000 independent samples in the MCMC chain for that model, which is a sufficient number to rule out C/O\,$>1$ at greater than $3\,\sigma$ confidence.  Second, the $\chi^2$ values for the best fit models provide additional evidence in favor of the O-rich scenario. The best fit O-rich model has $\chi^2 = 2.75$ versus $\chi^2 = 29.2$ for the best fit C-rich case (for 2 degrees of freedom).  This difference in $\chi^2$ implies the O-rich model is $10^6$ times more likely than the best fit C-rich model, assuming Gaussian statistics.  Third, the Bayes factor (the ratio of integrated posterior probabilities) for the C-rich scenario to the O-rich scenario is 14, which constitutes strong evidence in favor of the O-rich model \citep{jeffreys61}.  In addition, the fact that the temperature range drops unphysically low -- below the condensation temperature of water -- is further evidence that the carbon-rich model is not appropriate for these data.  

Taken together, these results rule out a carbon-rich composition for the atmosphere at high confidence.  This is a more definitive constraint than what we obtained from studying the water abundance alone because the model is sensitive to both the presence of water \emph{and} the absence of absorption features from other molecules.  For example, the fact that no methane features are detected in the spectrum strengthens the case for an O-rich composition beyond the constraints from the presence of water alone. The C-C model also assumes more prior knowledge by imposing chemical equilibrium in the calculation of the model spectra.

\section{Summary \& Conclusions}
\label{sec:conclusion}
We have measured a precise transmission spectrum for the hot Jupiter WASP-12b over the wavelength range 0.84 to 1.67\,$\mu$m with \textit{HST}/WFC3.  The transmission spectrum is a factor of three more precise than previous measurements in this wavelength range \citep{swain13}.  We retrieved the atmospheric properties based on this spectrum with a variety of models, including a new retrieval parameterization that fits for C/O and metallicity rather than molecular abundances. Our conclusions about the nature of the planet are summarized as follows: 
\begin{enumerate}
  \item{Water is present in the atmosphere.  Models that do not include water absorption are excluded at 7\,$\sigma$ confidence.  This result is the first unambiguous spectroscopic detection of a molecule in the planet's atmosphere.}
  \item{The 68\% credible interval for the retrieved water abundance is $10^{-5} - 10^{-2}$ for a wide range of models.  This result applies to scenarios where water is the only absorber included in the model, where clouds and haze are added, where additional absorbers are incorporated (CO, CO$_2$, NH$_3$, TiO, VO, Na, K, CH$_4$, C$_2$H$_2$, HCN, H$_2$S, FeH, and N$_2$), and where optical transit depth measurements from \textit{HST}/STIS are also fit.  This range agrees well with the predicted equilibrium water abundance for a solar composition atmosphere  but is still consistent to within 2$\sigma$ with the prediction for a carbon-rich composition.}  
  \item{Stellar photons are absorbed at temperatures of $1000-1900$ Kelvin and pressures of $0.1-10$ mbar (based on 1\,$\sigma$ ranges for $T_s$ and $P_\mathrm{eq}$ from the FULL model).  These estimates have large uncertainties because the transmission spectrum is only weakly sensitive to the thermal structure of the atmosphere.}
\item{Based on a new retrieval parameterization that fits for C/O and metallicity under the assumption of chemical equilibrium, we constrain the C/O to $0.5^{+0.2}_{-0.3}$ at $1\,\sigma$ and rule out a carbon-rich atmosphere composition (C/O\,$>1$) at $>3\,\sigma$ confidence.  With this model, we also detect the presence of clouds/haze at 3.7 $\sigma$ confidence.}
\end{enumerate}

Our constraint on C/O is in tension with past studies of the planet's atmosphere that found a carbon-rich composition was the best explanation for the dayside emission spectrum \citep[e.g.][]{madhusudhan11a, stevenson14b}.  We would hope for better agreement between these analyses, as both results are based on very high precision data fit with state-of-the-art retrieval models. This methodology has yielded consistent results for other planets, notably the hot Jupiter WASP-43b, which shows excellent agreement between estimates of the composition from the dayside emission and transmission spectra \citep{kreidberg14b}.

We note that a caveat for our results from the C-C model is our assumption that the atmosphere is in chemical equilibrium.  \cite{moses13} considered the effects of photochemistry and mixing for WASP-12b and found that the water abundance is either unchanged from equilibrium values (for solar composition), or pushed even lower (for a C/O = 1 composition). Based on these results, it is unlikely that our high observed water abundance is due to disequilibrium chemistry in a carbon-rich atmosphere.  However, these calculations were for the planet's dayside and it would be worth exploring disequilibrium effects for the terminator region specifically in future work.

Another assumption in our model is that the temperature structure is well approximated as 1D. In reality, a photon's slant path through the atmosphere traverses many different temperatures and pressures.  However, theoretical models predict that most of the stellar radiation is attenuated in a fairly localized region within a few degrees of the terminator \citep{fortney10}.  The change in temperature and pressure over a region this size is small compared to the uncertainty in our estimates of the thermal profile, so our assumption of a 1D profile is unlikely to bias our assessment of the atmospheric composition.   On the other hand, the observations integrate over the entire limb of the planet, and there could be large variations in temperature over this region due to atmospheric dynamics. It is therefore possible that fitting the limb-averaged spectrum with a single 1D thermal profile could bias the results.

One route to reconciling the disagreement between constraints on C/O from the transmission and emission spectra is to determine the planet's global thermal structure.  This could be achieved by combining phase-resolved emission spectroscopy of the planet with 3D atmospheric circulation modeling, as has been done for WASP-43b \citep{stevenson14c, kataria15}.  This combination would provide an independent measure of the temperature structure at the terminator to strengthen our interpretation of the transmission spectrum.  It would also put the dayside temperature-pressure profile in context of the planet's global heat circulation and aid in understanding the dayside emission spectrum. In addition, a spectroscopic phase curve would allow us to estimate the water abundance for new regions of the  planet's atmosphere.  Furthermore, the spectroscopic phase curve amplitudes themselves could provide an additional diagnostic of the atmosphere composition.  Day-night temperature differences for hot Jupiters are larger at lower pressures \citep{showman09, stevenson14c, kataria14, kataria15}, so light curves in absorption bands (which probe lower pressures) are expected to have larger amplitudes than those in spectral windows. Measuring the wavelength dependence of the phase curve amplitude could therefore provide an additional constraint on the atmospheric composition \citep{showman09, stevenson14c, kataria14, kataria15}.

Finally, our results highlight the necessity of obtaining high-precision data with multiple observing techniques (transmission spectroscopy, dayside emission spectroscopy, and phase curves) in order to obtain robust constraints on the rich chemistry and physics of exoplanet atmospheres.  Studying the atmosphere from more than one angle (literally) is key to providing a detailed understanding of its thermal structure and dynamics, which is needed to unambiguously determine the composition (and vice versa).  In addition, high-precision measurements are essential for revealing the small features present in exoplanet spectra.  The amplitude of these features has often been smaller than predicted for transiting planets, due to the presence of clouds or haze and shallower thermal profiles \citep[e.g.][]{fortney06,charbonneau02,deming13}.  Our WASP-12b spectrum is a new example of this for transmission measurements, with features crossing just two scale heights. We therefore advocate applying an intensive approach of high-precision spectroscopy from multiple angles to a larger sample of transiting exoplanets to shed light on their nature and origins.  Such measurements will help develop the observing strategies needed for definitive characterization of exoplanet atmospheres, and prepare the community to make robust measurements of potentially habitable worlds with future facilities.

\acknowledgments
This work is based on observations made with the NASA/ESA Hubble Space Telescope that were obtained at the Space Telescope Science Institute, which is operated by the Association of Universities for Research in Astronomy, Inc., under NASA contract NAS 5-26555.  These observations are associated with program GO-13467. Support for this work was provided by NASA through a grant from the Space Telescope Science Institute, the National Science Foundation through a Graduate Research Fellowship (to L.K.), the Alfred P. Sloan Foundation through a Sloan Research Fellowship (to J.L.B.), the Packard Foundation through a Packard Fellowship for Science and Engineering (to J.L.B.), and the Sagan Fellowship Program (to K.B.S.) as supported by NASA and administered by the NASA Exoplanet Science Institute.  G.W.H. and M.H.W. acknowledge support from Tennessee State University and the State of Tennessee through its Centers of Excellence program.

We thank Patrick Irwin for the use of his NEMESIS code.  We are grateful to Megan Bedell, Hannah Diamond-Lowe, and Greg Gilbert for helpful discussions about this work.  We thank Drake Deming, Peter McCullough, Adam Burrows, Sara Seager, Dave Charbonneau, and Derek Homeier for being co-investigators on the \textit{HST} observing proposal.

\bibliographystyle{apj}
\bibliography{ms.bib}

\begin{thebibliography}{}
\expandafter\ifx\csname natexlab\endcsname\relax\def\natexlab#1{#1}\fi

\bibitem[{{Ali-Dib} {et~al.}(2014){Ali-Dib}, {Mousis}, {Petit}, \&
  {Lunine}}]{alidib14}
{Ali-Dib}, M., {Mousis}, O., {Petit}, J.-M., \& {Lunine}, J.~I. 2014, \apj,
  785, 125

\bibitem[{{Asplund} {et~al.}(2009){Asplund}, {Grevesse}, {Sauval}, \&
  {Scott}}]{asplund09}
{Asplund}, M., {Grevesse}, N., {Sauval}, A.~J., \& {Scott}, P. 2009, \araa, 47,
  481

\bibitem[{{Atreya} {et~al.}(2003){Atreya}, {Mahaffy}, {Niemann}, {Wong}, \&
  {Owen}}]{atreya03}
{Atreya}, S.~K., {Mahaffy}, P.~R., {Niemann}, H.~B., {Wong}, M.~H., \& {Owen},
  T.~C. 2003, \planss, 51, 105

\bibitem[{{Atreya} {et~al.}(1999){Atreya}, {Wong}, {Owen}, {Mahaffy},
  {Niemann}, {de Pater}, {Drossart}, \& {Encrenaz}}]{atreya99}
{Atreya}, S.~K., {Wong}, M.~H., {Owen}, T.~C., {et~al.} 1999, \planss, 47, 1243

\bibitem[{{Barstow} {et~al.}(2013){Barstow}, {Aigrain}, {Irwin}, {Fletcher}, \&
  {Lee}}]{barstow13}
{Barstow}, J.~K., {Aigrain}, S., {Irwin}, P.~G.~J., {Fletcher}, L.~N., \&
  {Lee}, J.-M. 2013, \mnras, 434, 2616

\bibitem[{{Barstow} {et~al.}(2014){Barstow}, {Aigrain}, {Irwin}, {Hackler},
  {Fletcher}, {Lee}, \& {Gibson}}]{barstow14}
{Barstow}, J.~K., {Aigrain}, S., {Irwin}, P.~G.~J., {et~al.} 2014, \apj, 786,
  154

\bibitem[{{Bechter} {et~al.}(2014){Bechter}, {Crepp}, {Ngo}, {Knutson},
  {Batygin}, {Hinkley}, {Muirhead}, {Johnson}, {Howard}, {Montet}, {Matthews},
  \& {Morton}}]{bechter14}
{Bechter}, E.~B., {Crepp}, J.~R., {Ngo}, H., {et~al.} 2014, \apj, 788, 2

\bibitem[{{Benneke} \& {Seager}(2013)}]{benneke13}
{Benneke}, B., \& {Seager}, S. 2013, \apj, 778, 153

\bibitem[{{Bergfors} {et~al.}(2013){Bergfors}, {Brandner}, {Daemgen}, {Biller},
  {Hippler}, {Janson}, {Kudryavtseva}, {Gei{\ss}ler}, {Henning}, \&
  {K{\"o}hler}}]{bergfors13}
{Bergfors}, C., {Brandner}, W., {Daemgen}, S., {et~al.} 2013, \mnras, 428, 182

\bibitem[{{Berta} {et~al.}(2011){Berta}, {Charbonneau}, {Bean}, {Irwin},
  {Burke}, {D{\'e}sert}, {Nutzman}, \& {Falco}}]{berta11}
{Berta}, Z.~K., {Charbonneau}, D., {Bean}, J., {et~al.} 2011, \apj, 736, 12

\bibitem[{{Berta} {et~al.}(2012){Berta}, {Charbonneau}, {D{\'e}sert},
  {Miller-Ricci Kempton}, {McCullough}, {Burke}, {Fortney}, {Irwin}, {Nutzman},
  \& {Homeier}}]{berta12}
{Berta}, Z.~K., {Charbonneau}, D., {D{\'e}sert}, J.-M., {et~al.} 2012, \apj,
  747, 35

\bibitem[{{Brogi} {et~al.}(2014){Brogi}, {de Kok}, {Birkby}, {Schwarz}, \&
  {Snellen}}]{brogi14}
{Brogi}, M., {de Kok}, R.~J., {Birkby}, J.~L., {Schwarz}, H., \& {Snellen},
  I.~A.~G. 2014, \aap, 565, A124

\bibitem[{{Burrows} \& {Sharp}(1999)}]{burrows99}
{Burrows}, A., \& {Sharp}, C.~M. 1999, \apj, 512, 843

\bibitem[{{Burton} {et~al.}(2015){Burton}, {Watson}, {Rodr{\'{\i}}guez-Gil},
  {Skillen}, {Littlefair}, {Dhillon}, \& {Pollacco}}]{burton15}
{Burton}, J.~R., {Watson}, C.~A., {Rodr{\'{\i}}guez-Gil}, P., {et~al.} 2015,
  \mnras, 446, 1071

\bibitem[{{Campo} {et~al.}(2011){Campo}, {Harrington}, {Hardy}, {Stevenson},
  {Nymeyer}, {Ragozzine}, {Lust}, {Anderson}, {Collier-Cameron}, {Blecic},
  {Britt}, {Bowman}, {Wheatley}, {Loredo}, {Deming}, {Hebb}, {Hellier},
  {Maxted}, {Pollaco}, \& {West}}]{campo11}
{Campo}, C.~J., {Harrington}, J., {Hardy}, R.~A., {et~al.} 2011, \apj, 727, 125

\bibitem[{{Castelli} \& {Kurucz}(2004)}]{castelli04}
{Castelli}, F., \& {Kurucz}, R.~L. 2004, ArXiv Astrophysics e-prints,
  astro-ph/0405087

\bibitem[{{Charbonneau} {et~al.}(2002){Charbonneau}, {Brown}, {Noyes}, \&
  {Gilliland}}]{charbonneau02}
{Charbonneau}, D., {Brown}, T.~M., {Noyes}, R.~W., \& {Gilliland}, R.~L. 2002,
  \apj, 568, 377

\bibitem[{{Copperwheat} {et~al.}(2013){Copperwheat}, {Wheatley}, {Southworth},
  {Bento}, {Marsh}, {Dhillon}, {Fortney}, {Littlefair}, \&
  {Hickman}}]{copperwheat13}
{Copperwheat}, C.~M., {Wheatley}, P.~J., {Southworth}, J., {et~al.} 2013,
  \mnras, 434, 661

\bibitem[{{Cornish} \& {Littenberg}(2007)}]{cornish07}
{Cornish}, N.~J., \& {Littenberg}, T.~B. 2007, \prd, 76, 083006

\bibitem[{{Cowan} {et~al.}(2012){Cowan}, {Machalek}, {Croll}, {Shekhtman},
  {Burrows}, {Deming}, {Greene}, \& {Hora}}]{cowan12}
{Cowan}, N.~B., {Machalek}, P., {Croll}, B., {et~al.} 2012, \apj, 747, 82

\bibitem[{{Croll} {et~al.}(2011){Croll}, {Lafreniere}, {Albert},
  {Jayawardhana}, {Fortney}, \& {Murray}}]{croll11}
{Croll}, B., {Lafreniere}, D., {Albert}, L., {et~al.} 2011, \aj, 141, 30

\bibitem[{{Croll} {et~al.}(2015){Croll}, {Albert}, {Jayawardhana}, {Cushing},
  {Moutou}, {Lafreniere}, {Johnson}, {Bonomo}, {Deleuil}, \&
  {Fortney}}]{croll15}
{Croll}, B., {Albert}, L., {Jayawardhana}, R., {et~al.} 2015, \apj, 802, 28

\bibitem[{{Crossfield} {et~al.}(2012){Crossfield}, {Barman}, {Hansen},
  {Tanaka}, \& {Kodama}}]{crossfield12}
{Crossfield}, I.~J.~M., {Barman}, T., {Hansen}, B.~M.~S., {Tanaka}, I., \&
  {Kodama}, T. 2012, \apj, 760, 140

\bibitem[{{Deming} {et~al.}(2013){Deming}, {Wilkins}, {McCullough}, {Burrows},
  {Fortney}, {Agol}, {Dobbs-Dixon}, {Madhusudhan}, {Crouzet}, {Desert},
  {Gilliland}, {Haynes}, {Knutson}, {Line}, {Magic}, {Mandell}, {Ranjan},
  {Charbonneau}, {Clampin}, {Seager}, \& {Showman}}]{deming13}
{Deming}, D., {Wilkins}, A., {McCullough}, P., {et~al.} 2013, \apj, 774, 95

\bibitem[{{D{\'e}sert} {et~al.}(2011){D{\'e}sert}, {Sing}, {Vidal-Madjar},
  {H{\'e}brard}, {Ehrenreich}, {Lecavelier Des Etangs}, {Parmentier}, {Ferlet},
  \& {Henry}}]{desert11}
{D{\'e}sert}, J.-M., {Sing}, D., {Vidal-Madjar}, A., {et~al.} 2011, \aap, 526,
  A12

\bibitem[{{Diamond-Lowe} {et~al.}(2014){Diamond-Lowe}, {Stevenson}, {Bean},
  {Line}, \& {Fortney}}]{diamond-lowe14}
{Diamond-Lowe}, H., {Stevenson}, K.~B., {Bean}, J.~L., {Line}, M.~R., \&
  {Fortney}, J.~J. 2014, \apj, 796, 66

\bibitem[{{Dodson-Robinson} {et~al.}(2009){Dodson-Robinson}, {Willacy},
  {Bodenheimer}, {Turner}, \& {Beichman}}]{dodsonrobinson09}
{Dodson-Robinson}, S.~E., {Willacy}, K., {Bodenheimer}, P., {Turner}, N.~J., \&
  {Beichman}, C.~A. 2009, \icarus, 200, 672

\bibitem[{{F{\"o}hring} {et~al.}(2013){F{\"o}hring}, {Dhillon}, {Madhusudhan},
  {Marsh}, {Copperwheat}, {Littlefair}, \& {Wilson}}]{fohring13}
{F{\"o}hring}, D., {Dhillon}, V.~S., {Madhusudhan}, N., {et~al.} 2013, \mnras,
  435, 2268

\bibitem[{{Foreman-Mackey} {et~al.}(2013){Foreman-Mackey}, {Hogg}, {Lang}, \&
  {Goodman}}]{foremanmackey13}
{Foreman-Mackey}, D., {Hogg}, D.~W., {Lang}, D., \& {Goodman}, J. 2013, \pasp,
  125, 306

\bibitem[{{Fortney} {et~al.}(2006){Fortney}, {Cooper}, {Showman}, {Marley}, \&
  {Freedman}}]{fortney06}
{Fortney}, J.~J., {Cooper}, C.~S., {Showman}, A.~P., {Marley}, M.~S., \&
  {Freedman}, R.~S. 2006, \apj, 652, 746

\bibitem[{{Fortney} {et~al.}(2008){Fortney}, {Lodders}, {Marley}, \&
  {Freedman}}]{fortney08}
{Fortney}, J.~J., {Lodders}, K., {Marley}, M.~S., \& {Freedman}, R.~S. 2008,
  \apj, 678, 1419

\bibitem[{{Fortney} {et~al.}(2010){Fortney}, {Shabram}, {Showman}, {Lian},
  {Freedman}, {Marley}, \& {Lewis}}]{fortney10}
{Fortney}, J.~J., {Shabram}, M., {Showman}, A.~P., {et~al.} 2010, \apj, 709,
  1396

\bibitem[{{Fossati} {et~al.}(2010){Fossati}, {Haswell}, {Froning}, {Hebb},
  {Holmes}, {Kolb}, {Helling}, {Carter}, {Wheatley}, {Collier Cameron},
  {Loeillet}, {Pollacco}, {Street}, {Stempels}, {Simpson}, {Udry}, {Joshi},
  {West}, {Skillen}, \& {Wilson}}]{fossati10}
{Fossati}, L., {Haswell}, C.~A., {Froning}, C.~S., {et~al.} 2010, \apjl, 714,
  L222

\bibitem[{{Gautier} {et~al.}(2001){Gautier}, {Hersant}, {Mousis}, \&
  {Lunine}}]{gautier01}
{Gautier}, D., {Hersant}, F., {Mousis}, O., \& {Lunine}, J.~I. 2001, \apjl,
  550, L227

\bibitem[{{Guillot}(2010)}]{guillot10}
{Guillot}, T. 2010, \aap, 520, A27

\bibitem[{{Haswell} {et~al.}(2012){Haswell}, {Fossati}, {Ayres}, {France},
  {Froning}, {Holmes}, {Kolb}, {Busuttil}, {Street}, {Hebb}, {Collier Cameron},
  {Enoch}, {Burwitz}, {Rodriguez}, {West}, {Pollacco}, {Wheatley}, \&
  {Carter}}]{haswell12}
{Haswell}, C.~A., {Fossati}, L., {Ayres}, T., {et~al.} 2012, \apj, 760, 79

\bibitem[{{Hauschildt} {et~al.}(1999){Hauschildt}, {Allard}, \&
  {Baron}}]{hauschildt99}
{Hauschildt}, P.~H., {Allard}, F., \& {Baron}, E. 1999, \apj, 512, 377

\bibitem[{{Hebb} {et~al.}(2009){Hebb}, {Collier-Cameron}, {Loeillet},
  {Pollacco}, {H{\'e}brard}, {Street}, {Bouchy}, {Stempels}, {Moutou},
  {Simpson}, {Udry}, {Joshi}, {West}, {Skillen}, {Wilson}, {McDonald},
  {Gibson}, {Aigrain}, {Anderson}, {Benn}, {Christian}, {Enoch}, {Haswell},
  {Hellier}, {Horne}, {Irwin}, {Lister}, {Maxted}, {Mayor}, {Norton}, {Parley},
  {Pont}, {Queloz}, {Smalley}, \& {Wheatley}}]{hebb09}
{Hebb}, L., {Collier-Cameron}, A., {Loeillet}, B., {et~al.} 2009, \apj, 693,
  1920

\bibitem[{{Heng} {et~al.}(2012){Heng}, {Hayek}, {Pont}, \& {Sing}}]{heng12}
{Heng}, K., {Hayek}, W., {Pont}, F., \& {Sing}, D.~K. 2012, \mnras, 420, 20

\bibitem[{{Hersant} {et~al.}(2004){Hersant}, {Gautier}, \&
  {Lunine}}]{hersant04}
{Hersant}, F., {Gautier}, D., \& {Lunine}, J.~I. 2004, \planss, 52, 623

\bibitem[{{Irwin} {et~al.}(2008){Irwin}, {Teanby}, {de Kok}, {Fletcher},
  {Howett}, {Tsang}, {Wilson}, {Calcutt}, {Nixon}, \& {Parrish}}]{irwin08}
{Irwin}, P.~G.~J., {Teanby}, N.~A., {de Kok}, R., {et~al.} 2008, \jqsrt, 109,
  1136

\bibitem[{Jeffreys(1998)}]{jeffreys61}
Jeffreys, H. 1998, The Theory of Probability (OUP Oxford)

\bibitem[{Kass \& Raftery(1995)}]{kass95}
Kass, R.~E., \& Raftery, A.~E. 1995, Journal of the american statistical
  association, 90, 773

\bibitem[{{Kataria} {et~al.}(2014){Kataria}, {Showman}, {Fortney}, {Marley}, \&
  {Freedman}}]{kataria14}
{Kataria}, T., {Showman}, A.~P., {Fortney}, J.~J., {Marley}, M.~S., \&
  {Freedman}, R.~S. 2014, \apj, 785, 92

\bibitem[{{Kataria} {et~al.}(2015){Kataria}, {Showman}, {Fortney}, {Stevenson},
  {Line}, {Kreidberg}, {Bean}, \& {D{\'e}sert}}]{kataria15}
{Kataria}, T., {Showman}, A.~P., {Fortney}, J.~J., {et~al.} 2015, \apj, 801, 86

\bibitem[{{Knutson} {et~al.}(2007){Knutson}, {Charbonneau}, {Noyes}, {Brown},
  \& {Gilliland}}]{knutson07}
{Knutson}, H.~A., {Charbonneau}, D., {Noyes}, R.~W., {Brown}, T.~M., \&
  {Gilliland}, R.~L. 2007, \apj, 655, 564

\bibitem[{{Knutson} {et~al.}(2014){Knutson}, {Dragomir}, {Kreidberg},
  {Kempton}, {McCullough}, {Fortney}, {Bean}, {Gillon}, {Homeier}, \&
  {Howard}}]{knutson14b}
{Knutson}, H.~A., {Dragomir}, D., {Kreidberg}, L., {et~al.} 2014, \apj, 794,
  155

\bibitem[{{Konopacky} {et~al.}(2013){Konopacky}, {Barman}, {Macintosh}, \&
  {Marois}}]{konopacky13}
{Konopacky}, Q.~M., {Barman}, T.~S., {Macintosh}, B.~A., \& {Marois}, C. 2013,
  Science, 339, 1398

\bibitem[{{Kopparapu} {et~al.}(2012){Kopparapu}, {Kasting}, \&
  {Zahnle}}]{kopparapu12}
{Kopparapu}, R.~k., {Kasting}, J.~F., \& {Zahnle}, K.~J. 2012, \apj, 745, 77

\bibitem[{{Kreidberg} {et~al.}(2014{\natexlab{a}}){Kreidberg}, {Bean},
  {D{\'e}sert}, {Line}, {Fortney}, {Madhusudhan}, {Stevenson}, {Showman},
  {Charbonneau}, {McCullough}, {Seager}, {Burrows}, {Henry}, {Williamson},
  {Kataria}, \& {Homeier}}]{kreidberg14b}
{Kreidberg}, L., {Bean}, J.~L., {D{\'e}sert}, J.-M., {et~al.}
  2014{\natexlab{a}}, \apjl, 793, L27

\bibitem[{{Kreidberg} {et~al.}(2014{\natexlab{b}}){Kreidberg}, {Bean},
  {D{\'e}sert}, {Benneke}, {Deming}, {Stevenson}, {Seager}, {Berta-Thompson},
  {Seifahrt}, \& {Homeier}}]{kreidberg14a}
---. 2014{\natexlab{b}}, \nat, 505, 69

\bibitem[{{Lecavelier Des Etangs} {et~al.}(2008){Lecavelier Des Etangs},
  {Pont}, {Vidal-Madjar}, \& {Sing}}]{lecavelier08}
{Lecavelier Des Etangs}, A., {Pont}, F., {Vidal-Madjar}, A., \& {Sing}, D.
  2008, \aap, 481, L83

\bibitem[{{Line} {et~al.}(2013{\natexlab{a}}){Line}, {Knutson}, {Deming},
  {Wilkins}, \& {Desert}}]{line13b}
{Line}, M.~R., {Knutson}, H., {Deming}, D., {Wilkins}, A., \& {Desert}, J.-M.
  2013{\natexlab{a}}, \apj, 778, 183

\bibitem[{{Line} {et~al.}(2014){Line}, {Knutson}, {Wolf}, \& {Yung}}]{line14}
{Line}, M.~R., {Knutson}, H., {Wolf}, A.~S., \& {Yung}, Y.~L. 2014, \apj, 783,
  70

\bibitem[{{Line} {et~al.}(2011){Line}, {Vasisht}, {Chen}, {Angerhausen}, \&
  {Yung}}]{line11}
{Line}, M.~R., {Vasisht}, G., {Chen}, P., {Angerhausen}, D., \& {Yung}, Y.~L.
  2011, \apj, 738, 32

\bibitem[{{Line} {et~al.}(2013{\natexlab{b}}){Line}, {Wolf}, {Zhang},
  {Knutson}, {Kammer}, {Ellison}, {Deroo}, {Crisp}, \& {Yung}}]{line13a}
{Line}, M.~R., {Wolf}, A.~S., {Zhang}, X., {et~al.} 2013{\natexlab{b}}, \apj,
  775, 137

\bibitem[{{Lodders}(2004)}]{lodders04}
{Lodders}, K. 2004, \apj, 611, 587

\bibitem[{{L{\'o}pez-Morales} {et~al.}(2010){L{\'o}pez-Morales}, {Coughlin},
  {Sing}, {Burrows}, {Apai}, {Rogers}, {Spiegel}, \& {Adams}}]{lopez-morales10}
{L{\'o}pez-Morales}, M., {Coughlin}, J.~L., {Sing}, D.~K., {et~al.} 2010,
  \apjl, 716, L36

\bibitem[{{Madhusudhan}(2012)}]{madhusudhan12}
{Madhusudhan}, N. 2012, \apj, 758, 36

\bibitem[{{Madhusudhan} {et~al.}(2014{\natexlab{a}}){Madhusudhan}, {Amin}, \&
  {Kennedy}}]{madhusudhan14b}
{Madhusudhan}, N., {Amin}, M.~A., \& {Kennedy}, G.~M. 2014{\natexlab{a}},
  \apjl, 794, L12

\bibitem[{{Madhusudhan} {et~al.}(2014{\natexlab{b}}){Madhusudhan}, {Crouzet},
  {McCullough}, {Deming}, \& {Hedges}}]{madhusudhan14a}
{Madhusudhan}, N., {Crouzet}, N., {McCullough}, P.~R., {Deming}, D., \&
  {Hedges}, C. 2014{\natexlab{b}}, \apjl, 791, L9

\bibitem[{{Madhusudhan} {et~al.}(2011{\natexlab{a}}){Madhusudhan}, {Mousis},
  {Johnson}, \& {Lunine}}]{madhusudhan11b}
{Madhusudhan}, N., {Mousis}, O., {Johnson}, T.~V., \& {Lunine}, J.~I.
  2011{\natexlab{a}}, \apj, 743, 191

\bibitem[{{Madhusudhan} \& {Seager}(2009)}]{madhusudhan09}
{Madhusudhan}, N., \& {Seager}, S. 2009, \apj, 707, 24

\bibitem[{{Madhusudhan} {et~al.}(2011{\natexlab{b}}){Madhusudhan},
  {Harrington}, {Stevenson}, {Nymeyer}, {Campo}, {Wheatley}, {Deming},
  {Blecic}, {Hardy}, {Lust}, {Anderson}, {Collier-Cameron}, {Britt}, {Bowman},
  {Hebb}, {Hellier}, {Maxted}, {Pollacco}, \& {West}}]{madhusudhan11a}
{Madhusudhan}, N., {Harrington}, J., {Stevenson}, K.~B., {et~al.}
  2011{\natexlab{b}}, \nat, 469, 64

\bibitem[{{Mandel} \& {Agol}(2002)}]{mandel02}
{Mandel}, K., \& {Agol}, E. 2002, \apjl, 580, L171

\bibitem[{{Mandell} {et~al.}(2013){Mandell}, {Haynes}, {Sinukoff},
  {Madhusudhan}, {Burrows}, \& {Deming}}]{mandell13}
{Mandell}, A.~M., {Haynes}, K., {Sinukoff}, E., {et~al.} 2013, \apj, 779, 128

\bibitem[{McBride \& Gordon(1996)}]{mcbride96}
McBride, B.~J., \& Gordon, S. 1996, {Computer Program for Calculation of
  Complex Chemical Equilibrium Compositions and Applications II. User's Manual
  and Program Description}, Tech. rep., NASA

\bibitem[{{McCullough} \& {MacKenty}(2012)}]{mccullough12}
{McCullough}, P., \& {MacKenty}, J. 2012, {Considerations for using Spatial
  Scans with WFC3}, Tech. rep.

\bibitem[{{Moses} {et~al.}(2013){Moses}, {Madhusudhan}, {Visscher}, \&
  {Freedman}}]{moses13}
{Moses}, J.~I., {Madhusudhan}, N., {Visscher}, C., \& {Freedman}, R.~S. 2013,
  \apj, 763, 25

\bibitem[{{Moses} {et~al.}(2011){Moses}, {Visscher}, {Fortney}, {Showman},
  {Lewis}, {Griffith}, {Klippenstein}, {Shabram}, {Friedson}, {Marley}, \&
  {Freedman}}]{moses11}
{Moses}, J.~I., {Visscher}, C., {Fortney}, J.~J., {et~al.} 2011, \apj, 737, 15

\bibitem[{{Nichols} {et~al.}(2015){Nichols}, {Wynn}, {Goad}, {Alexander},
  {Casewell}, {Cowley}, {Burleigh}, {Clarke}, \& {Bisikalo}}]{nichols15}
{Nichols}, J.~D., {Wynn}, G.~A., {Goad}, M., {et~al.} 2015, ArXiv e-prints,
  arXiv:1502.07489

\bibitem[{{{\"O}berg} {et~al.}(2011){{\"O}berg}, {Murray-Clay}, \&
  {Bergin}}]{oberg11}
{{\"O}berg}, K.~I., {Murray-Clay}, R., \& {Bergin}, E.~A. 2011, \apjl, 743, L16

\bibitem[{{Owen} {et~al.}(1999){Owen}, {Mahaffy}, {Niemann}, {Atreya},
  {Donahue}, {Bar-Nun}, \& {de Pater}}]{owen99}
{Owen}, T., {Mahaffy}, P., {Niemann}, H.~B., {et~al.} 1999, \nat, 402, 269

\bibitem[{{Parmentier} \& {Guillot}(2014)}]{parmentier14}
{Parmentier}, V., \& {Guillot}, T. 2014, \aap, 562, A133

\bibitem[{{Robinson} \& {Catling}(2012)}]{robinson12}
{Robinson}, T.~D., \& {Catling}, D.~C. 2012, \apj, 757, 104

\bibitem[{Sellke {et~al.}(2001)Sellke, Bayarri, \& Berger}]{sellke01}
Sellke, T., Bayarri, M.~J., \& Berger, J.~O. 2001, The American Statistician,
  55, pp. 62

\bibitem[{{Showman} {et~al.}(2009){Showman}, {Fortney}, {Lian}, {Marley},
  {Freedman}, {Knutson}, \& {Charbonneau}}]{showman09}
{Showman}, A.~P., {Fortney}, J.~J., {Lian}, Y., {et~al.} 2009, \apj, 699, 564

\bibitem[{{Sing} {et~al.}(2013){Sing}, {Lecavelier des Etangs}, {Fortney},
  {Burrows}, {Pont}, {Wakeford}, {Ballester}, {Nikolov}, {Henry}, {Aigrain},
  {Deming}, {Evans}, {Gibson}, {Huitson}, {Knutson}, {Showman}, {Vidal-Madjar},
  {Wilson}, {Williamson}, \& {Zahnle}}]{sing13}
{Sing}, D.~K., {Lecavelier des Etangs}, A., {Fortney}, J.~J., {et~al.} 2013,
  \mnras, 436, 2956

\bibitem[{{Stevenson} {et~al.}(2014{\natexlab{a}}){Stevenson}, {Bean},
  {Fabrycky}, \& {Kreidberg}}]{stevenson14d}
{Stevenson}, K.~B., {Bean}, J.~L., {Fabrycky}, D., \& {Kreidberg}, L.
  2014{\natexlab{a}}, \apj, 796, 32

\bibitem[{{Stevenson} {et~al.}(2014{\natexlab{b}}){Stevenson}, {Bean},
  {Madhusudhan}, \& {Harrington}}]{stevenson14b}
{Stevenson}, K.~B., {Bean}, J.~L., {Madhusudhan}, N., \& {Harrington}, J.
  2014{\natexlab{b}}, \apj, 791, 36

\bibitem[{{Stevenson} {et~al.}(2014{\natexlab{c}}){Stevenson}, {Bean},
  {Seifahrt}, {D{\'e}sert}, {Madhusudhan}, {Bergmann}, {Kreidberg}, \&
  {Homeier}}]{stevenson14a}
{Stevenson}, K.~B., {Bean}, J.~L., {Seifahrt}, A., {et~al.} 2014{\natexlab{c}},
  \aj, 147, 161

\bibitem[{{Stevenson} {et~al.}(2014{\natexlab{d}}){Stevenson}, {D{\'e}sert},
  {Line}, {Bean}, {Fortney}, {Showman}, {Kataria}, {Kreidberg}, {McCullough},
  {Henry}, {Charbonneau}, {Burrows}, {Seager}, {Madhusudhan}, {Williamson}, \&
  {Homeier}}]{stevenson14c}
{Stevenson}, K.~B., {D{\'e}sert}, J.-M., {Line}, M.~R., {et~al.}
  2014{\natexlab{d}}, Science, 346, 838

\bibitem[{{Swain} {et~al.}(2013){Swain}, {Deroo}, {Tinetti}, {Hollis},
  {Tessenyi}, {Line}, {Kawahara}, {Fujii}, {Showman}, \& {Yurchenko}}]{swain13}
{Swain}, M., {Deroo}, P., {Tinetti}, G., {et~al.} 2013, \icarus, 225, 432

\bibitem[{{Swain} {et~al.}(2014){Swain}, {Line}, \& {Deroo}}]{swain14}
{Swain}, M.~R., {Line}, M.~R., \& {Deroo}, P. 2014, \apj, 784, 133

\bibitem[{{Teske} {et~al.}(2014){Teske}, {Cunha}, {Smith}, {Schuler}, \&
  {Griffith}}]{teske14}
{Teske}, J.~K., {Cunha}, K., {Smith}, V.~V., {Schuler}, S.~C., \& {Griffith},
  C.~A. 2014, \apj, 788, 39

\bibitem[{{Teske} {et~al.}(2013){Teske}, {Schuler}, {Cunha}, {Smith}, \&
  {Griffith}}]{teske13}
{Teske}, J.~K., {Schuler}, S.~C., {Cunha}, K., {Smith}, V.~V., \& {Griffith},
  C.~A. 2013, \apjl, 768, L12

\bibitem[{{Trotta}(2008)}]{trotta08}
{Trotta}, R. 2008, Contemporary Physics, 49, 71

\bibitem[{{Venot} {et~al.}(2015){Venot}, {H{\'e}brard}, {Ag{\'u}ndez}, {Decin},
  \& {Bounaceur}}]{venot15}
{Venot}, O., {H{\'e}brard}, E., {Ag{\'u}ndez}, M., {Decin}, L., \& {Bounaceur},
  R. 2015, ArXiv e-prints, arXiv:1502.03567

\bibitem[{Weinberg(2012)}]{weinberg12}
Weinberg, M.~D. 2012, Bayesian Anal., 7, 737

\bibitem[{{Zhao} {et~al.}(2012){Zhao}, {Monnier}, {Swain}, {Barman}, \&
  {Hinkley}}]{zhao12}
{Zhao}, M., {Monnier}, J.~D., {Swain}, M.~R., {Barman}, T., \& {Hinkley}, S.
  2012, \apj, 744, 122

\end{thebibliography}

\end{document}